\documentclass[namedreferences]{solarphysics}
\usepackage[hyperref,optionalrh,solaromanenum]{spr-sola-addons}
\usepackage{graphicx}
\usepackage{color}
\usepackage{breakurl}

\usepackage{solaheader}
\newcommand{\solareceiveddate}{23 January 2015}
\newcommand{\solaaccepteddate}{29 August 2015}
\newcommand{\solaonlinedate}{18 September 2015}
\newcommand{\doi}{015-0770-4}

\renewcommand{\TODAY}{3 November 2009}

\newcommand{\grad}{\mbox{\boldmath$\nabla$}}
\newcommand{\vdot}{{\mathbf{\cdot}}}
\newcommand{\B}{\mathbf{B}}

\newcommand{\bxi}{B_x^{\rm i}}
\newcommand{\byi}{B_y^{\rm i}}
\newcommand{\bzi}{B_z^{\rm i}}
\newcommand{\bxh}{B_x^{\rm h}}
\newcommand{\byh}{B_y^{\rm h}}
\newcommand{\bzh}{B_z^{\rm h}}

\newcommand{\xuh}{x^{\rm h}}
\newcommand{\yuh}{y^{\rm h}}
\newcommand{\zuh}{z^{\rm h}}
\newcommand{\xui}{x^{\rm i}}
\newcommand{\yui}{y^{\rm i}}
\newcommand{\zui}{z^{\rm i}}
\newcommand{\myie}{{\it i.e., }}
\newcommand{\myeg}{{\it e.g., }}
\newcommand{\cbl}{{Paper~I}}
\newcommand{\adc}{{Paper~II}}
\newcommand{\st}{S} 
\newcommand{\ls}{\lambda_{\rm s}} 
\newcommand{\hs}{h_{\rm s}}
\newcommand{\hc}{h_{\rm c}}

\newcommand{\aap}{    {\it Astron. Astrophys.}}

\newcommand{\apj}{    {\it Astrophys. J.}}
\newcommand{\apjl}{    {\it Astrophys. J. Lett.}}
\newcommand{\apjs}{    {\it Astrophys. J. Suppl.}}

\newcommand{\jcp}{    {\it J. Chem. Phys.}}

\newcommand{\solphys}{{\it Solar Phys.}}

\begin{document}
\begin{article}
\begin{opening}

\title{Resolving the Azimuthal Ambiguity in Vector Magnetogram Data with the Divergence-Free Condition: Implementations for Disambiguating Each Height Independently}

\author{A.D.~\surname{Crouch}}

\runningauthor{A.D.~Crouch}
\runningtitle{Resolving the Azimuthal Ambiguity in Vector Magnetogram Data}

\institute{A.D.~Crouch\\NorthWest Research Associates, 3380 Mitchell Lane, Boulder, CO 80301, USA\\email: \url{ash@nwra.com}}

\begin{abstract}
We continue the investigation of how to use the divergence-free condition to resolve the azimuthal ambiguity present in vector magnetogram data.
In previous articles, 
by \citeauthor{2009SoPh..260..271C}~(\textit{Solar Phys.} {\bf 260}, 271, \citeyear{2009SoPh..260..271C})
and \citeauthor{2012SoPh..tmp..249C}~(\textit{Solar Phys.} {\bf 282}, 107, \citeyear{2012SoPh..tmp..249C}),
all  methods used an expression for the divergence of the magnetic field that involves differentiation of quantities that depend on the choice of azimuthal angle.
As a result, all heights used to approximate line-of-sight derivatives should generally be disambiguated simultaneously.
In this article, we investigate a set of methods that use an expression for the divergence that involves differentiation of quantities that do not depend on the choice of azimuthal angle.
This results in an expression for the divergence that can be used to disambiguate each height independently.
We test two methods using synthetic and find that the two-step, hybrid method, adapted to disambiguate each height independently, generally produces reasonable results.
Moreover, the time required to compute solutions is  substantially decreased in comparison to the corresponding method that disambiguates all relevant heights simultaneously.
\end{abstract}


\end{opening}

\section{Introduction}

Reliable measurements of the physical conditions in the solar atmosphere, such as the magnetic field, are required to constrain models for a wide range of solar phenomena.
To fully determine the magnetic field vector, however, involves the resolution of the azimuthal ambiguity, which
arises when the component of the field in the direction perpendicular to the line-of-sight is inferred from the linear polarisation of magnetically sensitive spectral lines (\myeg{} \opencite{1969PhDT.........3H}). 
This approach cannot distinguish between the two possible choices for the perpendicular component of the field, which have the same magnitude but differ in direction by 180$^\circ$.

Several methods are currently in use for resolving the azimuthal ambiguity in single-height vector magnetogram data; a selection of these were reviewed and tested on synthetic data by \inlinecite{2006SoPh..237..267M} and \inlinecite{2009SoPh..260...83L}.
Single-height vector magnetogram data include information about the variation of the magnetic field over the solar surface, but not in the direction perpendicular to it.
Therefore, disambiguation algorithms that need this information (say, to compute the divergence of the field) must make an assumption.
For example, some versions of the minimum energy method use a potential-field extrapolation to approximate the variation of the magnetic field in the direction perpendicular to the solar surface, which is needed to compute the divergence (\myeg{} \opencite{1994SoPh..155..235M}; \opencite{2006SoPh..237..267M}; \opencite{2009ASPC..415..365L}; \opencite{2009SoPh..260...83L}).
This approximation may not be an accurate representation of the solar magnetic field in general.

Another class of methods, that are currently under development, use multiple-height vector magnetogram data together with the divergence-free condition to resolve the azimuthal ambiguity (\myeg{} \opencite{1990AcApS..10..371W}; \opencite{1993A+A...278..279C}; \opencite{1993A+A...279..214L}; \opencite{1999A+A...347.1005B}; \opencite{2007ApJ...654..675L}; \opencite{2008SoPh..247...25C}; \opencite{2009SoPh..260..271C}; \opencite{2012SoPh..tmp..249C}).
Multiple-height vector magnetogram data include information about the variation of the magnetic field along the line-of-sight,
which can be used to  approximate the derivatives of the magnetic field in the line-of-sight direction.
Subsequently, this information can be used to compute the divergence of the magnetic field for any position on the solar disk (for details see \opencite{2008SoPh..247...25C}; \opencite{2009SoPh..260..271C}; \opencite{2012SoPh..tmp..249C}).
However, there are several methods available that each make a different assumption about how to employ the divergence-free condition to resolve the ambiguity.
The differences in the implementation details can have a significant effect on the results produced; for comparisons and tests using synthetic data see \citeauthor{2009SoPh..260..271C}~(\citeyear{2009SoPh..260..271C}, henceforth \cbl{}) and \citeauthor{2012SoPh..tmp..249C}~(\citeyear{2012SoPh..tmp..249C}, henceforth \adc{}).

All of the methods examined in \cbl{} and \adc{} use an expression for the divergence of the field that involves differentiation of quantities that depend on the choice of azimuthal angle.
As a result, all heights used to approximate line-of-sight derivatives should generally be disambiguated simultaneously.
In contrast, the purpose of this article is to investigate methods that use an expression for the divergence that involves differentiation of quantities that do not depend on the choice of azimuthal angle (\myeg{} \opencite{1996SoPh..164..291S}; \opencite{1998A+A...331..383S}; \opencite{2003JKAS...36S...7S}).
As we will show, this leads to an expression for the divergence that can be used to disambiguate each height independently.
This expression is derived in Section~\ref{sec_divb}.
In Section~\ref{sec_synth} we review the synthetic data and metrics that are  used to test the performance of the various disambiguation methods.
In Section~\ref{sec_wuai} we re-examine the \inlinecite{1990AcApS..10..371W} criterion for resolving the azimuthal ambiguity using the expression for the divergence derived in Section~\ref{sec_divb}.
In Section~\ref{sec_hybrid} we test the hybrid method that was presented in \adc{}, adapted to use the expression for the divergence derived in Section~\ref{sec_divb}.
In Section~\ref{sec_conc} we present conclusions.

\section{A Modified Expression For the Divergence-Free Condition}
\label{sec_divb}

Following \inlinecite{2008SoPh..247...25C}, \cbl{}, and \adc{}, in this section we briefly re-iterate the derivation of the expression for the divergence of the magnetic field in terms of observable quantities that is valid for any position on the solar disk.
We then modify the expression obtained for the divergence by replacing derivatives of quantities that are dependent on the choice of azimuthal angle with derivatives of quantities that are independent of the choice of azimuthal angle, using an approach  similar to \inlinecite{1996SoPh..164..291S}, \inlinecite{1998A+A...331..383S}, and \inlinecite{2003JKAS...36S...7S}.
Subsequently, we discuss the consequences for using the modified version of the divergence-free condition to resolve the azimuthal ambiguity in vector magnetogram data.

Throughout this investigation we consider multiple-height, vector-magnetogram data within a field of view that is limited in spatial extent such that a layer of constant optical depth can be approximated by the  $\xuh$\,--\,$\yuh$-plane (where the superscript h denotes heliographic coordinates).
At each pixel at each observation height, we assume that the image components of the magnetic field vector can be measured (denoted with the superscript i):
\(\bxi\),
\(\byi\), and
\(\bzi\),
where
\(\bzi = B_\|\) is the line-of-sight component
and

\begin{equation}
\bxi = B_\perp \cos \xi \, , \qquad \mbox{and} \qquad \byi = B_\perp \sin \xi \, ,
\label{btrans}
\end{equation}

\noindent
where \(B_\perp\) is the magnitude of the transverse component (perpendicular to the line-of-sight) and \(\xi\) is the azimuthal angle.
It is important to emphasise that, when using the linear polarisation of magnetically sensitive spectral lines, the azimuthal angle \(\xi\)  can only be inferred within the range \(0 \leq \xi < 180^\circ\), which results in the azimuthal ambiguity (\myie it is not possible to distinguish between the two choices of azimuthal angle, \(\xi\) and \(\xi + 180^\circ\), without additional information).
The relationship between the heliographic and image components of the field is

\begin{eqnarray}
\bxh & = & a_{11} \bxi + a_{12} \byi + a_{13} \bzi \, , \nonumber \\
\byh & = & a_{21} \bxi + a_{22} \byi + a_{23} \bzi \, , \label{B_h} \\
\bzh & = & a_{31} \bxi + a_{32} \byi + a_{33} \bzi \, ,  \nonumber
\end{eqnarray}

\noindent
where the coefficients $a_{ij}$ are taken from Equation~(1) of \inlinecite{1990SoPh..126...21G} and are assumed to be constant within the field of view.

When the divergence of the magnetic field is expressed in heliographic coordinates, the derivative of the field in the direction perpendicular to the $\xuh$\,--\,$\yuh$-plane is required (\myeg \(\partial \bzh / \partial \zuh\)), but this derivative cannot be directly inferred from observations except at disk centre.
This is because the available methods for inferring the variation of the magnetic field vector out of the $\xuh$\,--\,$\yuh$-plane can do so only in the line-of-sight direction (\myeg \opencite{1992ApJ...398..375R}; \opencite{1994A+A...291..622C}; \opencite{1995ApJ...439..474M}; \opencite{1996SoPh..164..169D}; \opencite{1996SoPh..169...79L}; \opencite{1998ApJ...494..453W}, \citeyear{2001ApJ...547.1130W}; \opencite{2000ApJ...530..977S}; \opencite{2002A+A...381..290E}; \opencite{2003SoPh..212..361L}; \opencite{2005ApJ...631L.167S}, \citeyear{2007ApJS..169..439S}).
For convenience, we assume that the variation of the field can be inferred as a function of line-of-sight distance \(\zui\), although we acknowledge that in practice the line-of-sight variation of the field may be inferred as a function of optical depth \(\tau\); in such cases, the relationship between \(\tau\) and \(\zui\) can be determined using a solar atmosphere model (\myeg \opencite{1986ApJ...306..284M}; \opencite{1981ApJS...45..635V}; \opencite{1994A+A...291..622C}; \opencite{2007ApJS..169..439S}).
Using Equations~(\ref{B_h}), it can be shown that the relationship between derivatives with respect to \(\zuh\) and \(\zui\) is

\begin{equation}
\frac{\partial f}{\partial \zui} = a_{13} \frac{\partial f}{ \partial \xuh} + a_{23} \frac{\partial f}{ \partial \yuh} + a_{33} \frac{\partial f}{\partial \zuh} \, ,
\label{dlos}
\end{equation}

\noindent
where $f$ is any differentiable function.
In practice, all of the derivatives required to compute the divergence must be approximated because the field is measured at discrete locations: in the horizontal heliographic directions this is because the field is sampled discretely over the heliographic plane (\myeg in pixels), whereas, in the line-of-sight direction this is because the magnetic field is generally inferred at discrete heights (see \citeauthor{2008SoPh..247...25C}~(\citeyear{2008SoPh..247...25C}) and \cbl{} for a summary of the different methods for inferring the line-of-sight variation of the magnetic field).

Starting with the divergence of the magnetic field expressed in heliographic coordinates, Equations~(\ref{B_h}) and (\ref{dlos}) can be used to rewrite the expression in terms of observable quantities (\myie derivatives of the image components of the field with respect to $\xuh$, $\yuh$ and $\zui$), which is valid for any position on the solar disk,

\begin{equation}
a_{33} \grad \vdot \B  = D_a + a_{33} \frac{\partial  \bzi}{\partial \zui} \, , 
\label{divb2}
\end{equation}

\noindent
where

\begin{eqnarray}
D_a & = & a_{31} \frac{\partial \bxi}{\partial \zui} + a_{32} \frac{\partial \byi}{\partial \zui}
+ \left( a_{11} a_{33} - a_{13} a_{31} \right) \frac{\partial \bxi}{\partial \xuh}
+ \left( a_{12} a_{33} - a_{13} a_{32} \right) \frac{\partial \byi}{\partial \xuh} \nonumber \\
& + & \left( a_{21} a_{33} - a_{23} a_{31} \right) \frac{\partial \bxi}{\partial \yuh}
+ \left( a_{22} a_{33} - a_{23} a_{32} \right) \frac{\partial \byi}{\partial \yuh} \, . \label{da}
\end{eqnarray}

\noindent
Each of the terms in \(D_a\) (Equation~(\ref{da})) involve differentiation of quantities that are sensitive to the choice of azimuthal angle (\myie \(\bxi\) and \(\byi\)).
For example, in the line-of-sight direction, this means that vector magnetogram data should be disambiguated simultaneously at all heights used to approximate line-of-sight derivatives.
\cbl{} and \adc{} examined several methods  based on Equations~(\ref{divb2}) and (\ref{da}).
The purpose of this article is to investigate methods that are based on a different expression.
By using an approach similar to that of \inlinecite{1996SoPh..164..291S}, \inlinecite{1998A+A...331..383S}, and \inlinecite{2003JKAS...36S...7S}, we  modify this expression for the divergence so that differentiation only operates on quantities that are independent of the choice of azimuthal angle, such as \(\cos^2 \xi\), \(\sin^2 \xi\), \(\cos \xi \sin \xi\), \( \cos \left( 2 \xi \right) \), and \( \sin \left( 2 \xi \right) \); we prefer to work with these quantities instead of \((\bxi)^2\), \((\byi)^2\), and \( \bxi \byi \) because methods based on these perform slightly better in tests (results not shown).
For example, we can use

\begin{displaymath}
\frac{\partial \cos^2 \xi}{\partial \xuh} = 2 \cos \xi  \frac{\partial \cos \xi}{\partial \xuh} \, ,
\end{displaymath}

\begin{displaymath}
\frac{\partial \sin^2 \xi}{\partial \xuh} = 2 \sin \xi  \frac{\partial \sin \xi}{\partial \xuh} \, ,
\end{displaymath}

\noindent
and

\begin{displaymath}
\frac{\partial  \cos \xi \sin \xi}{\partial \xuh} = \sin \xi  \frac{\partial \cos \xi}{\partial \xuh} +  \cos \xi  \frac{\partial \sin \xi}{\partial \xuh} \, ,
\end{displaymath}

\noindent
to show that

\begin{eqnarray}
  2 \frac{\partial \cos \xi}{\partial \xuh} & = & \cos \xi \frac{ \partial \cos^2 \xi }{\partial \xuh} + 2 \sin \xi \frac{ \partial \cos \xi \sin \xi}{\partial \xuh} - \cos \xi \frac{ \partial \sin^2 \xi }{\partial \xuh} \label{dcdxa} \\
  & = & \cos \xi \frac{ \partial \cos \left( 2 \xi \right)}{\partial \xuh} + \sin \xi \frac{ \partial \sin \left( 2 \xi \right)}{\partial \xuh}  \label{dcdxb} \, ,
\end{eqnarray}

\noindent
and

\begin{eqnarray}
  2 \frac{\partial \sin \xi}{\partial \xuh} & = & \sin \xi \frac{ \partial \sin^2 \xi }{\partial \xuh} + 2 \cos \xi \frac{ \partial \cos \xi \sin \xi}{\partial \xuh} - \sin \xi \frac{ \partial \cos^2 \xi }{\partial \xuh} \label{dsdxa} \\
  & = & \cos \xi \frac{ \partial \sin \left( 2 \xi \right)}{\partial \xuh} - \sin \xi \frac{ \partial \cos \left( 2 \xi \right)}{\partial \xuh} \label{dsdxb} \, .
\end{eqnarray}

\noindent
For all of the terms on the right hand sides of Equations~(\ref{dcdxa})~--~(\ref{dsdxb}), differentiation only operates on quantities that are independent of the choice of azimuthal angle.
Subsequently, Equations~(\ref{dcdxb}) and (\ref{dsdxb}) can be used to show that:

\begin{eqnarray}
2 \frac{\partial \bxi}{\partial \xuh} & = & 2 B_\perp \frac{\partial \cos \xi }{\partial \xuh} + 2 \cos \xi \frac{\partial B_\perp }{\partial \xuh} \nonumber \\
& = & \cos \xi \left[ 2 \frac{ \partial B_\perp}{\partial \xuh} + B_\perp \frac{ \partial \cos \left( 2 \xi \right) }{\partial \xuh} \right]
+  B_\perp \sin \xi \frac{ \partial \sin \left( 2 \xi \right) }{\partial \xuh} \, ,
\label{dbxdx}
\end{eqnarray}

\noindent
and

\begin{eqnarray}
2 \frac{\partial \byi}{\partial \xuh} & = & 2 B_\perp \frac{\partial \sin \xi }{\partial \xuh} + 2 \sin \xi \frac{\partial B_\perp }{\partial \xuh} \nonumber \\
& = & \sin \xi \left[ 2 \frac{ \partial B_\perp}{\partial \xuh} - B_\perp \frac{ \partial \cos \left( 2 \xi \right) }{\partial \xuh} \right]
+ B_\perp \cos \xi \frac{ \partial \sin \left( 2 \xi \right) }{\partial \xuh}  \, . 
\label{dbydx}
\end{eqnarray}

\noindent
Again, for all of the terms on the right hand sides of Equations~(\ref{dbxdx}) and (\ref{dbydx}), differentiation only operates on quantities that are independent of the choice of azimuthal angle.
Analogous expressions can be derived for the derivatives of \(\bxi\) and \(\byi\) with respect to both \(\yuh\) and \(\zui\).
Using Equations~(\ref{dbxdx}) and (\ref{dbydx}) and their counterparts, we can re-write the divergence-free condition as

\begin{equation}
2 a_{33} \grad \vdot \B  = \hs \sin \xi  + \hc \cos \xi + 2 a_{33} \frac{\partial  \bzi}{\partial \zui} \, , 
\label{divbafd}
\end{equation}

\noindent
where

\begin{eqnarray}
\hs & = & a_{31} B_\perp \frac{ \partial \sin \left( 2 \xi \right) }{\partial \zui}
+ a_{32} \left[ 2 \frac{\partial B_\perp}{\partial \zui} - B_\perp \frac{\partial \cos \left( 2 \xi \right) }{\partial \zui} \right]
\nonumber\\
& + & \left( a_{11} a_{33} - a_{13} a_{31} \right) B_\perp \frac{\partial \sin \left( 2 \xi \right) }{\partial \xuh}
\nonumber\\
& + & \left( a_{12} a_{33} - a_{13} a_{32} \right) \left[ 2 \frac{\partial B_\perp}{\partial \xuh} - B_\perp \frac{\partial \cos \left( 2 \xi \right) }{\partial \xuh} \right]
\nonumber\\
& + & \left( a_{21} a_{33} - a_{23} a_{31} \right) B_\perp \frac{\partial \sin \left( 2 \xi \right) }{\partial \yuh}
\nonumber\\
& + & \left( a_{22} a_{33} - a_{23} a_{32} \right) \left[ 2 \frac{\partial B_\perp}{\partial \yuh} - B_\perp \frac{\partial \cos \left( 2 \xi \right) }{\partial \yuh} \right] \, ,
\label{hseqn}
\end{eqnarray}

\noindent
and

\begin{eqnarray}
\hc & = &  a_{31} \left[ 2 \frac{\partial B_\perp}{\partial \zui} + B_\perp \frac{\partial \cos \left( 2 \xi \right) }{\partial \zui} \right]
+ a_{32} B_\perp \frac{ \partial \sin \left( 2 \xi \right) }{\partial \zui}
\nonumber\\
& + &  \left( a_{11} a_{33} - a_{13} a_{31} \right) \left[ 2 \frac{\partial B_\perp}{\partial \xuh} + B_\perp \frac{\partial \cos \left( 2 \xi \right) }{\partial \xuh} \right]
\nonumber\\
& + & \left( a_{12} a_{33} - a_{13} a_{32} \right) B_\perp \frac{\partial \sin \left( 2 \xi \right) }{\partial \xuh}
\nonumber\\
& + & \left( a_{21} a_{33} - a_{23} a_{31} \right)  \left[ 2 \frac{\partial B_\perp}{\partial \yuh} + B_\perp \frac{\partial \cos \left( 2 \xi \right) }{\partial \yuh} \right]
\nonumber\\
& + & \left( a_{22} a_{33} - a_{23} a_{32} \right) B_\perp \frac{\partial \sin \left( 2 \xi \right) }{\partial \yuh} \, .
\label{hceqn}
\end{eqnarray}

\noindent
The quantities \(\hs\) and \(\hc\) are independent of the choice of azimuthal angle.
Moreover, in \(\hs\) and \(\hc\), differentiation only operates on quantities that are independent of the choice of azimuthal angle.
This means that the approximation for the divergence of the magnetic field at one location does not need to depend on the choice of azimuthal angles in neighbouring locations (if  the various quantities are evaluated at the centre of each pixel at each height).
Therefore, it is possible to develop disambiguation methods based on Equations~(\ref{divbafd})~--~(\ref{hceqn}) such that the choice of azimuthal angle at one location does not depend on the choice of azimuthal angle at other locations (in any direction).
This can have a significant impact on methods for resolving the azimuthal ambiguity with the divergence-free condition.
For example, in the line-of-sight direction, using a method based on Equations~(\ref{divbafd})~--~(\ref{hceqn}), each height can be disambiguated independently; this is different to the methods examined in \cbl{} and \adc{}, which were based on Equations~(\ref{divb2}) and (\ref{da}).
It should be noted that line-of-sight derivatives of \( \cos \left( 2 \xi \right) \),  \( \sin \left( 2 \xi \right) \), \( B_\perp \), and \( \bzi \) are generally required to compute the divergence using Equations~(\ref{divbafd})~--~(\ref{hceqn}), but these do not depend on the choice of azimuthal angle.

\section{Synthetic Data and Performance Metrics}
\label{sec_synth}

To test the performance of the various algorithms based on the divergence-free condition, as described by Equations~(\ref{divbafd})~--~(\ref{hceqn}), we use the same synthetic data that were used in \cbl{} and \adc{} for which the correct configuration of azimuthal angles is known and measurements of the magnetic field are available at two heights.
These synthetic datasets are based on the same magnetic field configurations that were used by \inlinecite{2006SoPh..237..267M} and \inlinecite{2009SoPh..260...83L} and, therefore, allow direct comparisons with the  single-height methods  tested in those articles.
Three classes of synthetic data are considered:
\begin{enumerate}
\item
A snapshot from the simulations of \inlinecite{fan04} of a twisted magnetic flux tube emerging into an overlying potential field arcade.
This magnetogram is situated at disk centre and does not include the effects of noise.
Further details can be found in \inlinecite{2006SoPh..237..267M} and \cbl{}.
This synthetic dataset  was provided by Yuhong Fan.
\item
A multipole, linear-force free magnetic field constructed from a set of point sources located on a plane below the solar surface, located away from disk centre, centred at ($9^\circ$S, $36^\circ$E).
Three cases are constructed with different levels of noise added to the polarisation spectra: one no-noise case and two noise-added cases.
Further details are provided in \inlinecite{2009SoPh..260...83L} and \adc{}.
These synthetic datasets  were provided by K.D. Leka and Graham Barnes.
\item
A potential magnetic field with fine-scale structure with properties similar to penumbrae and plage, located at disk centre, and sampled on a grid with a pixel size of \(0.03''\).
To simulate the effects of imperfect instrumental spatial resolution three magnetograms are constructed with the polarisation spectra spatially binned by factors of 5, 10, and 30, resulting in effective pixel sizes of \(0.15'', 0.3''\), and \(0.9''\), respectively.
Some of the fine-scale structures that are fully resolved at \(0.03''\) are not fully resolved at lower spatial resolution.
For some additional discussion of this test case see \inlinecite{2009SoPh..260...83L}, \inlinecite{2012SoPh..277...89L}, \inlinecite{2012SoPh..276..423G}, \inlinecite{2012SoPh..276..441L} and \adc{}.
These synthetic datasets  were also provided by K.D. Leka and Graham Barnes.
\end{enumerate}

We emphasise that obtaining measurements of the magnetic field at two heights for solar observational data is a challenge that is beyond the scope of this investigation.
In principle, the line-of-sight variation of the magnetic field can be inferred from observations (\myeg \opencite{1992ApJ...398..375R}; \opencite{1994A+A...291..622C}; \opencite{1995ApJ...439..474M}; \opencite{1996SoPh..164..169D}; \opencite{1996SoPh..169...79L}; \opencite{1998ApJ...494..453W}, \citeyear{2001ApJ...547.1130W}; \opencite{2000ApJ...530..977S}; \opencite{2002A+A...381..290E}; \opencite{2003SoPh..212..361L}; \opencite{2005ApJ...631L.167S}, \citeyear{2007ApJS..169..439S}).
However, further research is required to develop methods for reliably obtaining maps of the solar magnetic field vector at multiple heights; this is especially true if one of the observation heights is located in the chromosphere (where interpretation difficulties are involved).

There are other sources of uncertainty that are not modelled by the synthetic data used in this investigation that may be significant for solar observational data:
i) it is assumed that the location of the observation height is known exactly (\myie the effects of limited spatial resolution in the line-of-sight direction are ignored);
ii) the magnetic field is assumed to be measured at constant geometrical height (constant optical depth may be more appropriate); and 
iii) the physical distance between the two observation heights is assumed to be constant (but this may vary over the field of view).
It is important to test the influence of these sources of uncertainty before the methods presented in this article are applied to solar observational data.

To measure the performance of the algorithms we use:
\( \mathcal{M}_{\rm area} \), the fraction of pixels correctly disambiguated, 
and \( \mathcal{M}_{B_\perp  > \mathcal{T}} \), the fraction of transverse magnetic field correctly disambiguated above a threshold \( \mathcal{T} \); we present results for \( \mathcal{T} = 100\)~G (gauss) and \( \mathcal{T} = 500\)~G.

\section{The Wu and Ai (1990) Criterion}
\label{sec_wuai}

The idea behind the \inlinecite{1990AcApS..10..371W} criterion for resolving the azimuthal ambiguity is to express the equation for the divergence-free condition as an inequality (\myeg \opencite{1993A+A...278..279C}; \opencite{1993A+A...279..214L}; \opencite{2007ApJ...654..675L}; \opencite{2008SoPh..247...25C}; \cbl{}; \adc{}).
For example, assuming \(\grad \vdot \B = 0\),  Equation~(\ref{divbafd}) can be used to show that

\begin{equation}
a_{33} \frac{\partial \bzi}{\partial \zui} \left( \hs \sin \xi  + \hc \cos \xi \right) = - 2 \left( a_{33} \frac{\partial  \bzi}{\partial \zui} \right)^2  \leq 0 \, .
\label{wuaieqn}
\end{equation}

\noindent
Thus, given the sign of \(\partial \bzi / \partial \zui\), Equation~(\ref{wuaieqn}) may be used to resolve the ambiguity, by choosing the realisation of the azimuthal angle that produces the correct sign for the quantity \( \hs \sin \xi  + \hc \cos \xi  \).
The magnitude of \(\partial \bzi / \partial \zui\) is not required; however, regarding various line-of-sight derivatives, it should be noted that the sign and magnitude of the line-of-sight derivatives of \( B_\perp \), \( \cos \left( 2 \xi \right)  \), and \( \sin \left( 2 \xi \right) \) are required to implement Equation~(\ref{wuaieqn}), except for locations where \(a_{31}=0\) and \(a_{32}=0\).

We implement the \inlinecite{1990AcApS..10..371W} criterion using Equation~(\ref{wuaieqn}) as follows.
Hereafter, pixel labels in the \(\xui\)-, \(\yui\)-, and \(\zui\)-directions are given by \(i\), \(j\), and \(k\) respectively, and the number of pixels in the \(\xui\)- and \(\yui\)-directions are given by \(n_x\) and \(n_y\), respectively.
For pixels with \(1 \leq i < n_x\) and \(1 \leq j < n_y\), the horizontal heliographic derivatives \(\partial / \partial \xuh\) and \(\partial / \partial \yuh\) at pixel \((i,j,k)\) are approximated with three-point finite differences using measurements of the magnetic field at pixels: \((i,j,k)\), \((i+1,j,k)\), and \((i,j+1,k)\).
At the boundaries where $i=n_x$ and $j=n_y$, the approximations for the horizontal heliographic derivatives are modified to use only pixels within the field of view.
At pixel \((i,j,k)\) line-of-sight derivatives \(\partial / \partial \zui\) are approximated with forward finite differences using measurements of the magnetic field from two heights, at pixels \((i,j,k)\) and \((i,j,k+1)\).
Subsequently, at each pixel the various terms in Equation~(\ref{wuaieqn}) are computed.
If the inequality is satisfied  the choice of azimuthal angle is not changed.
Otherwise, the choice of azimuthal angle is switched.
Three potentially advantageous features of this approach are i) the disambiguation result at one location does not depend on the choice of azimuthal angles in neighbouring locations; ii) it is non-iterative, meaning that each pixel only needs to be visited once during disambiguation; and iii) it is non-sequential, in that the results do not depend on the order in which pixels are visited.

Some limitations of the \inlinecite{1990AcApS..10..371W} criterion implemented with Equations~(\ref{divb2}) and (\ref{da}) were demonstrated in \cbl{}. 
These include
i) the disambiguation results depend on the smoothness of the initial configuration of azimuthal angles;
ii) the disambiguation results depend on the order in which pixels are visited; and
iii) the approximation for \(D_a\) in Equation~(\ref{da}) can have the same sign for both the correct and the incorrect choices of azimuthal angle (when derivatives are approximated from discrete measurements).
The implementation of the \inlinecite{1990AcApS..10..371W} criterion based on Equation~(\ref{wuaieqn}), as described above, does not suffer from these problems.
This is because the derivatives in \(\hs\) and \(\hc\) (see Equations~(\ref{hseqn}) and (\ref{hceqn}), respectively) only depend on quantities that are independent of the choice of azimuthal angle.
Furthermore, because \(\hs\) and \(\hc\) are independent of the choice of azimuthal angle, the quantity \( \hs \sin \xi + \hc \cos \xi \) changes only in sign (not magnitude) when the choice of azimuthal angle is switched at a given pixel at a given height.
However, the divergence of the magnetic field is generally not exactly zero when derivatives are approximated from discrete measurements and, therefore, some erroneous solutions are expected.

\begin{figure}[ht]
\begin{center}
\begin{tabular}{c@{\hspace{0.005\textwidth}}c@{\hspace{0.005\textwidth}}c}
\includegraphics[width=0.3\textwidth]{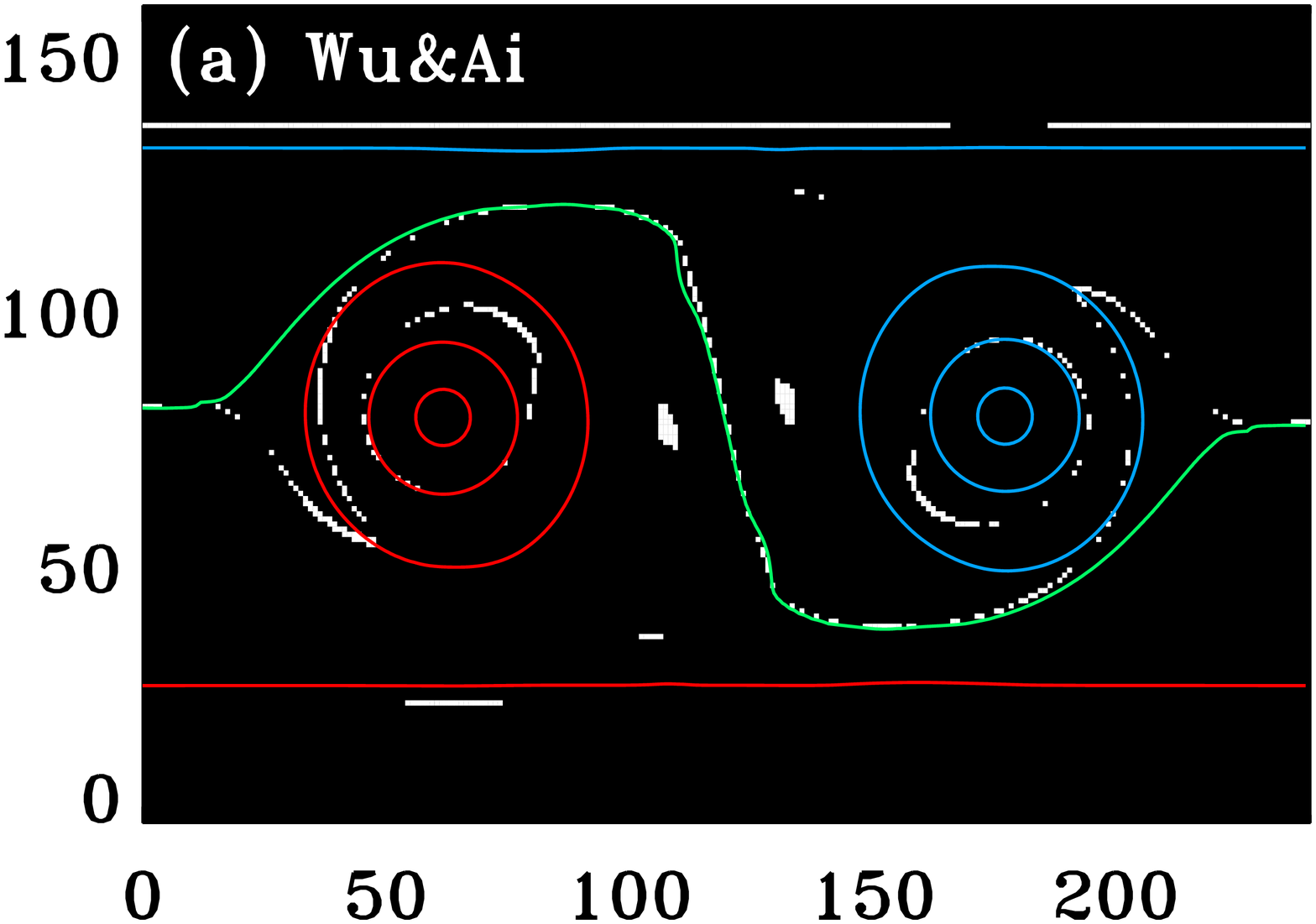}  &
\includegraphics[width=0.3\textwidth]{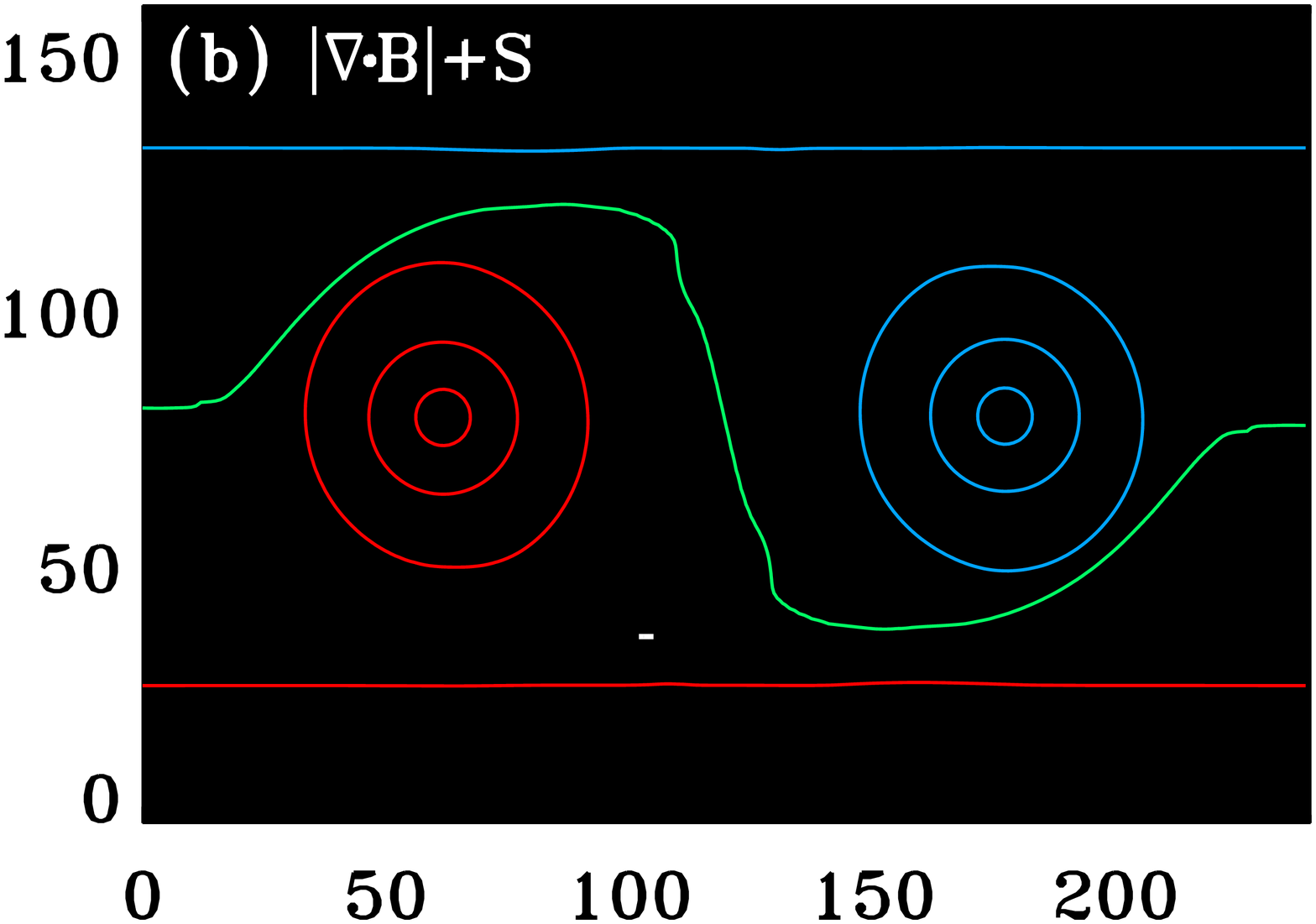}  &
\includegraphics[width=0.3\textwidth]{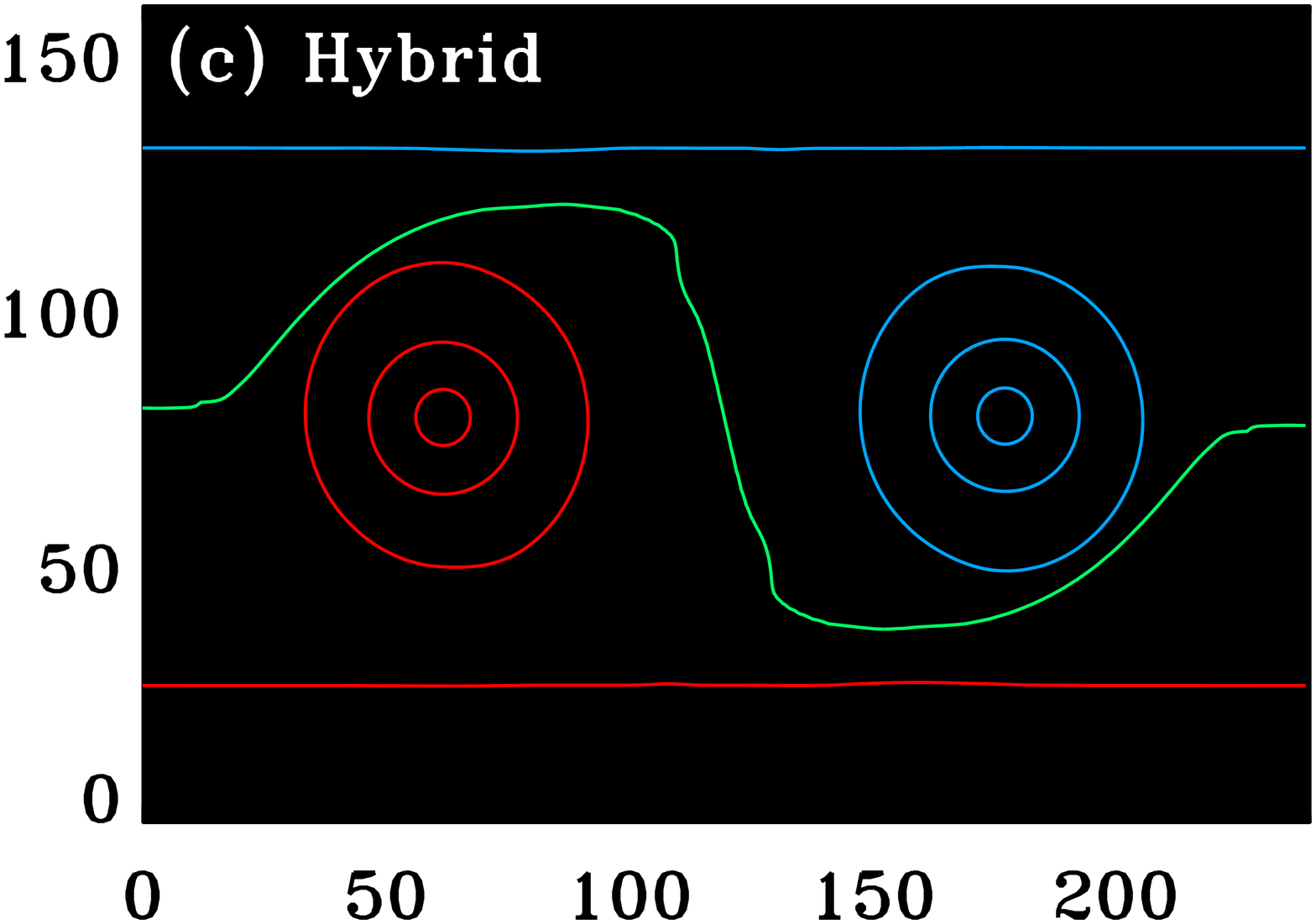}
\end{tabular}
\end{center}
\caption{Results for the various methods applied to the twisted flux tube and arcade.
Areas with the correct/incorrect azimuthal angle are black/white.
Positive/negative vertical magnetic flux is indicated by red/blue contours at 100, 1000, 2000~gauss~(G).
The magnetic neutral line is indicated by the green contour.
(a) Results from the Wu and Ai (1990) criterion.
(b) Results from the global minimisation method, which minimises \(E_{\rm s}\) (see Equation~(\ref{eds3})).
(c) Results from the hybrid method, which first minimises \(E_{\rm s}\) as in (b) (see Equation~(\ref{eds3})), then applies a smoothing algorithm to pixels below a threshold transverse field strength (as described in Section~\ref{sec_hybrid}).
}
\label{fig_fan56}
\end{figure}

\begin{figure}[ht]
\begin{center}
\begin{tabular}{c@{\hspace{0.005\textwidth}}c@{\hspace{0.005\textwidth}}c}
\includegraphics[width=0.3\textwidth]{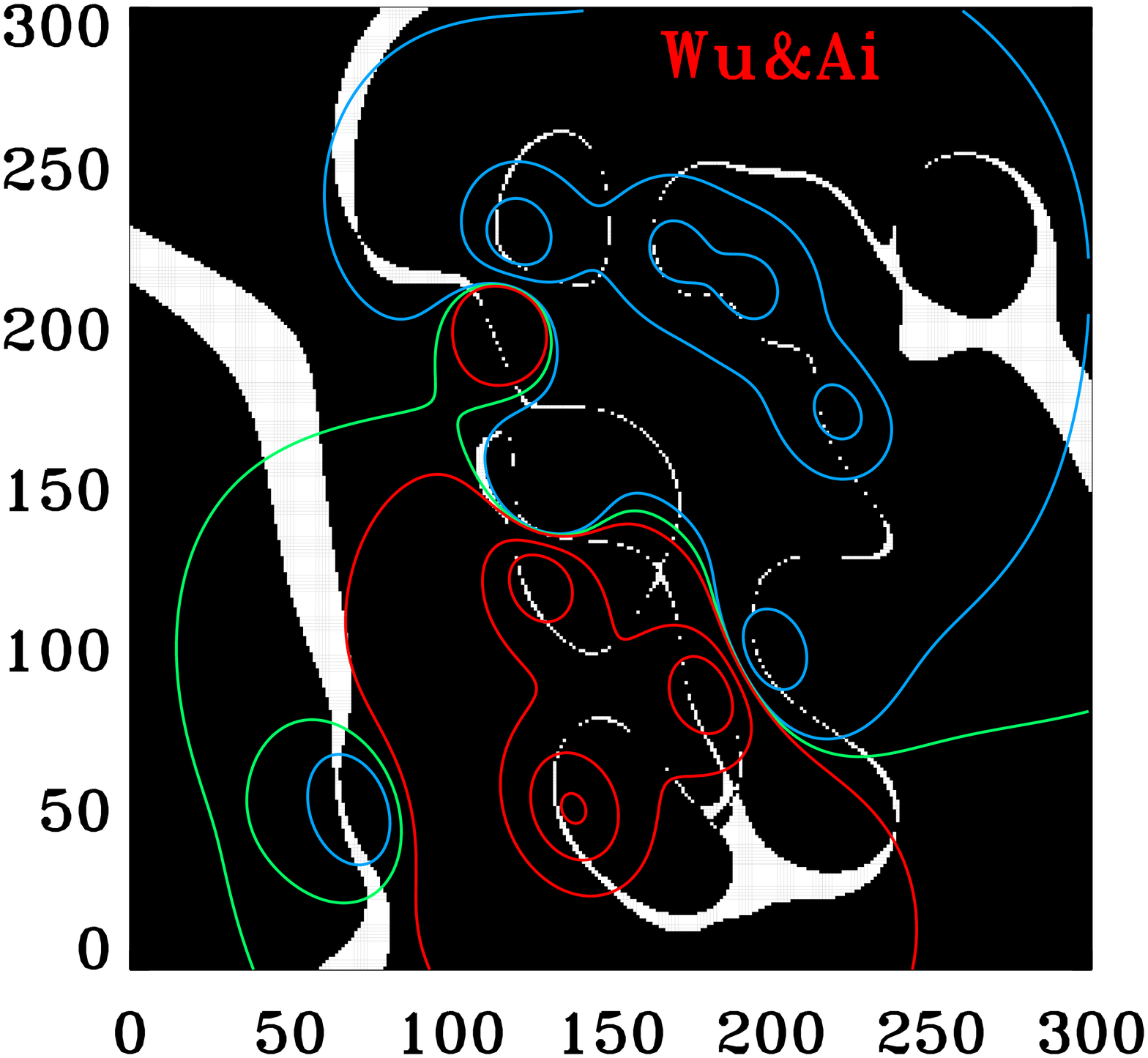} &
\includegraphics[width=0.3\textwidth]{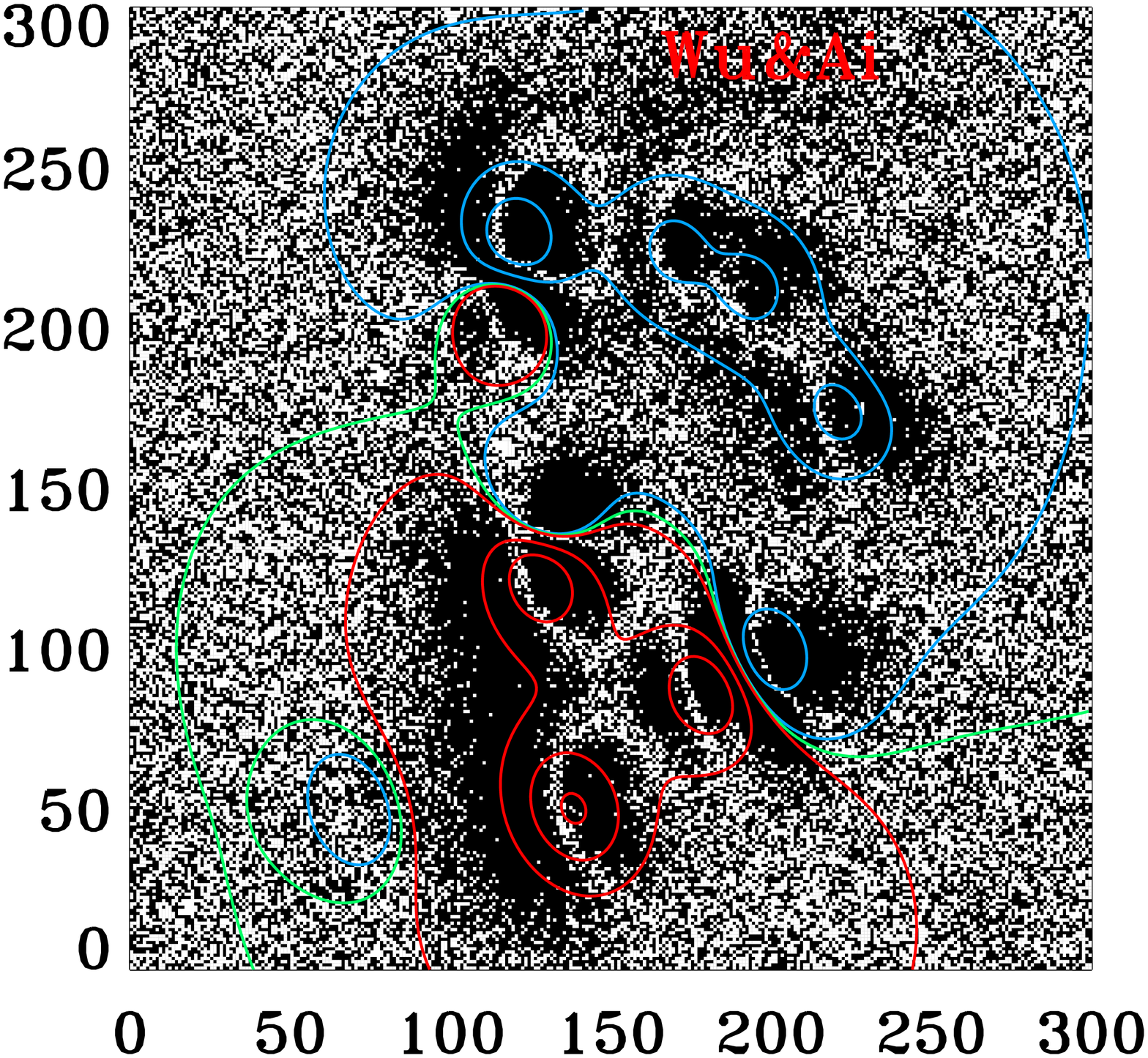} &
\includegraphics[width=0.3\textwidth]{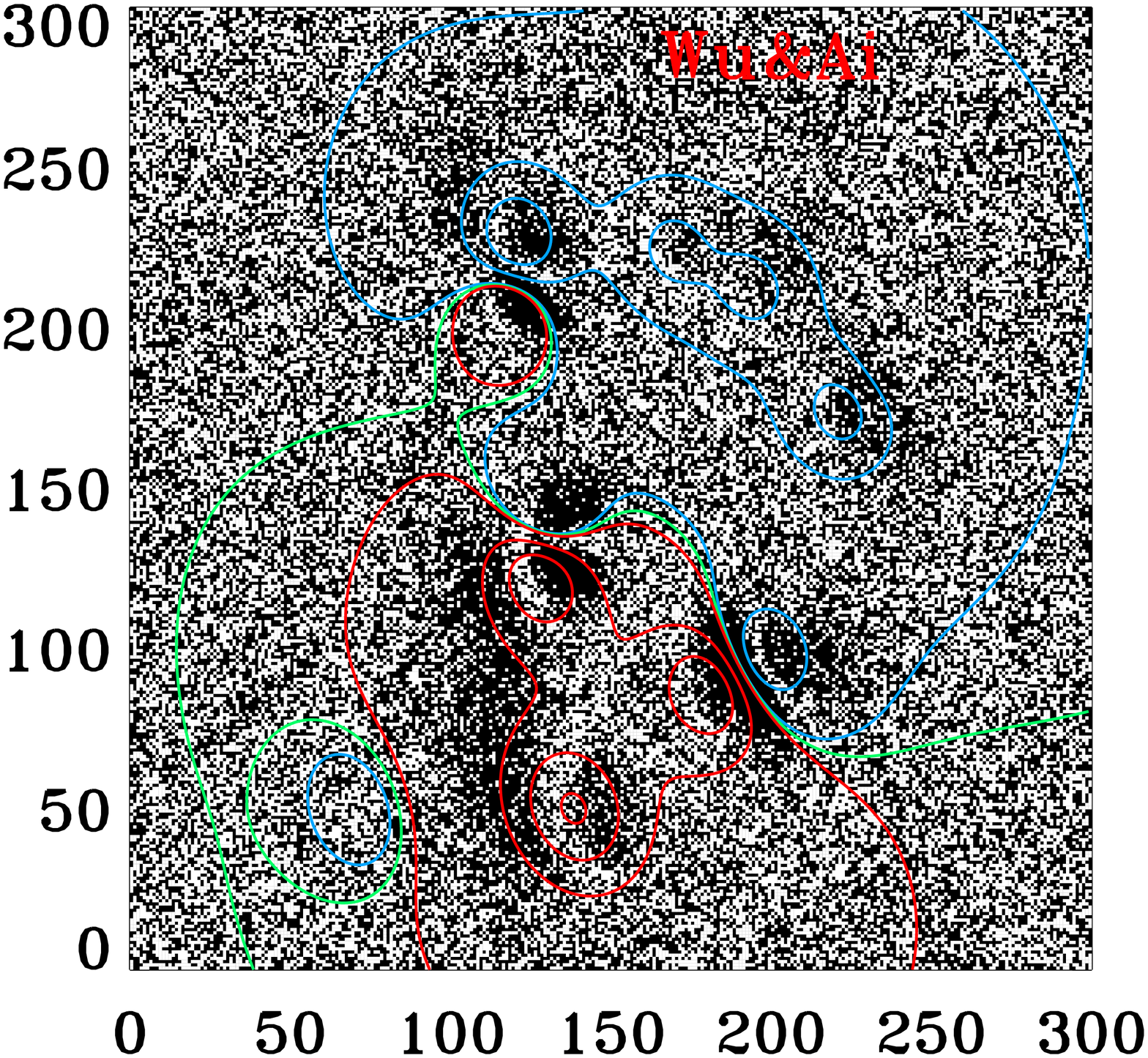} \\
\includegraphics[width=0.3\textwidth]{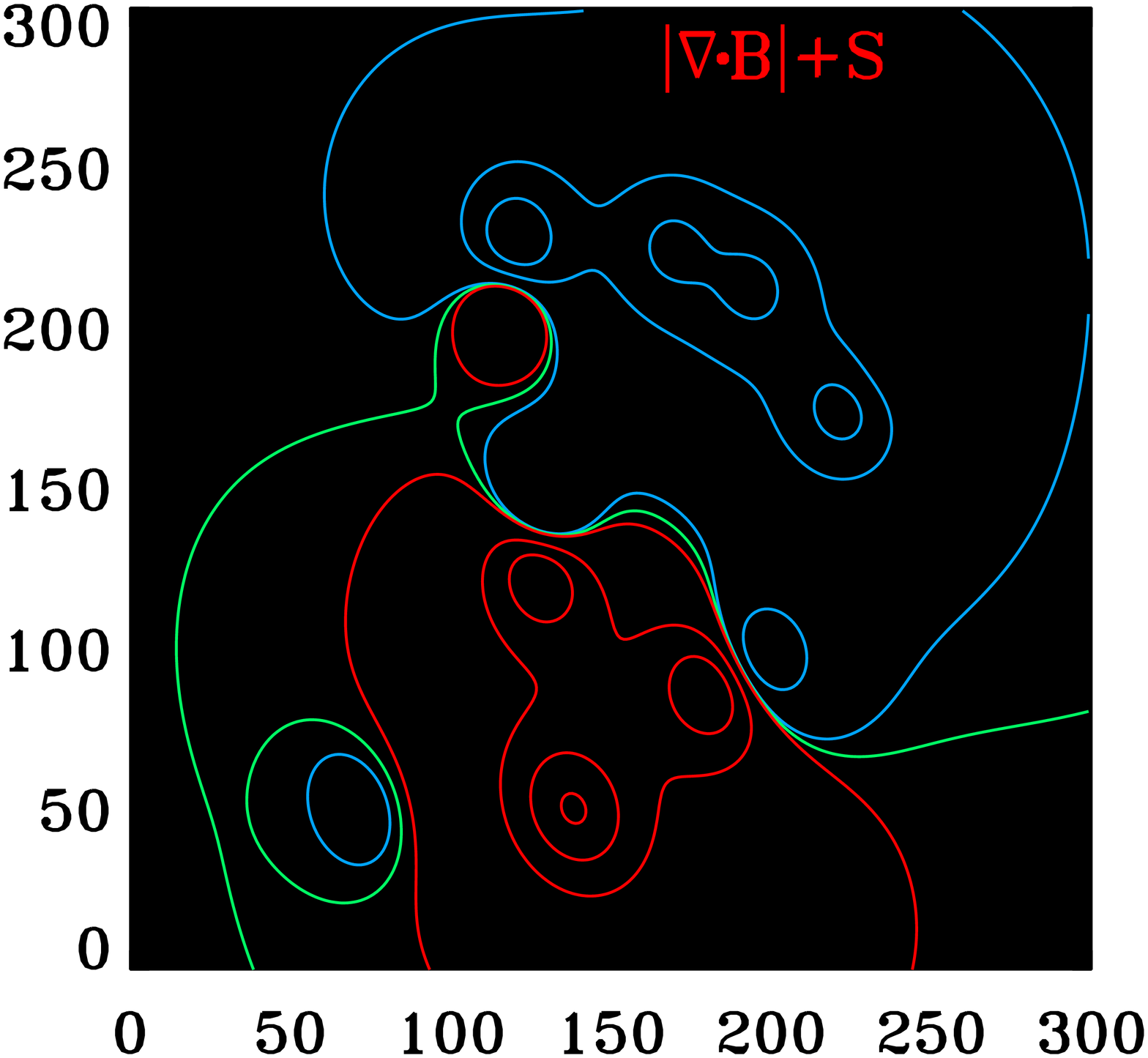} &
\includegraphics[width=0.3\textwidth]{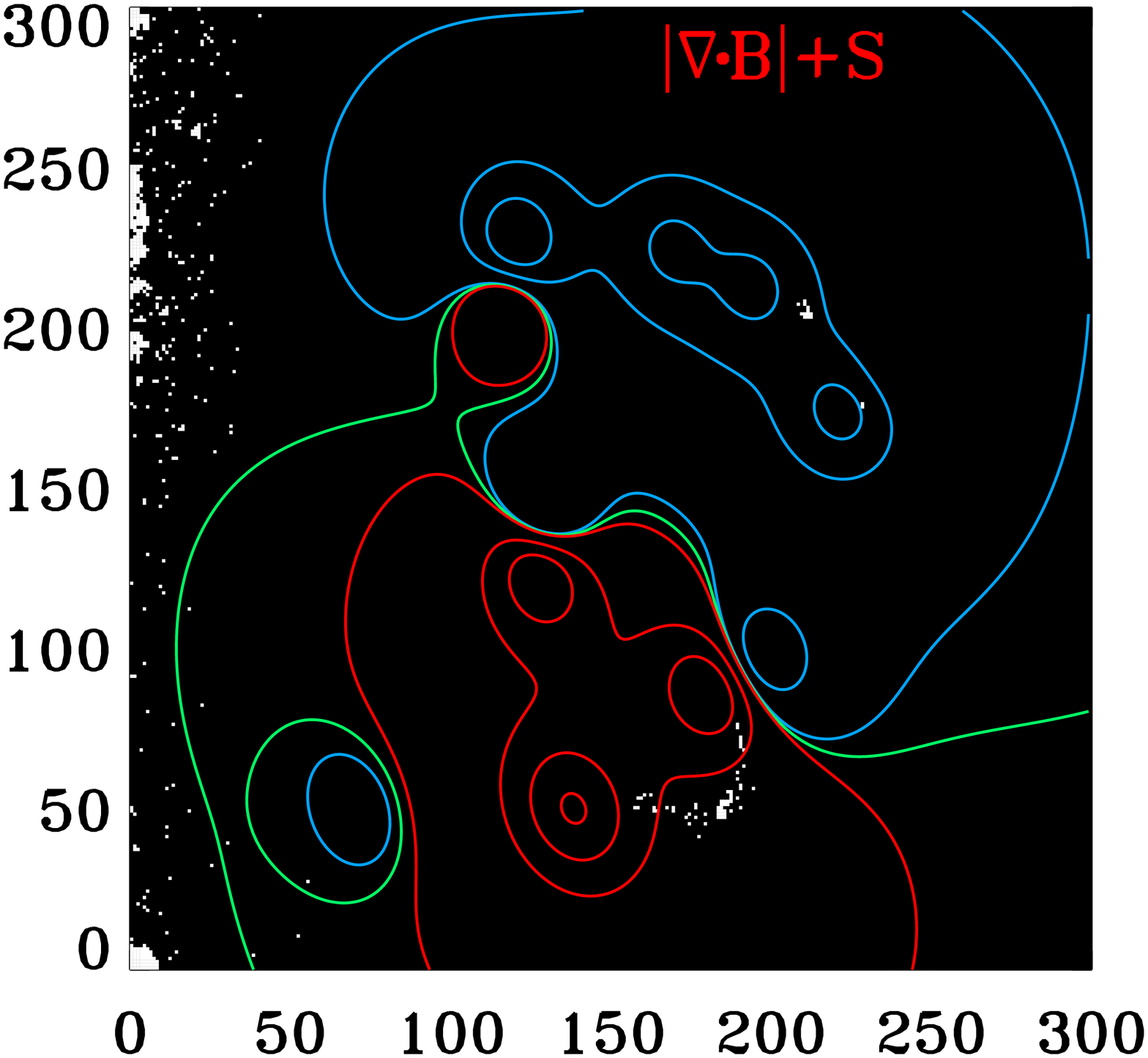} &
\includegraphics[width=0.3\textwidth]{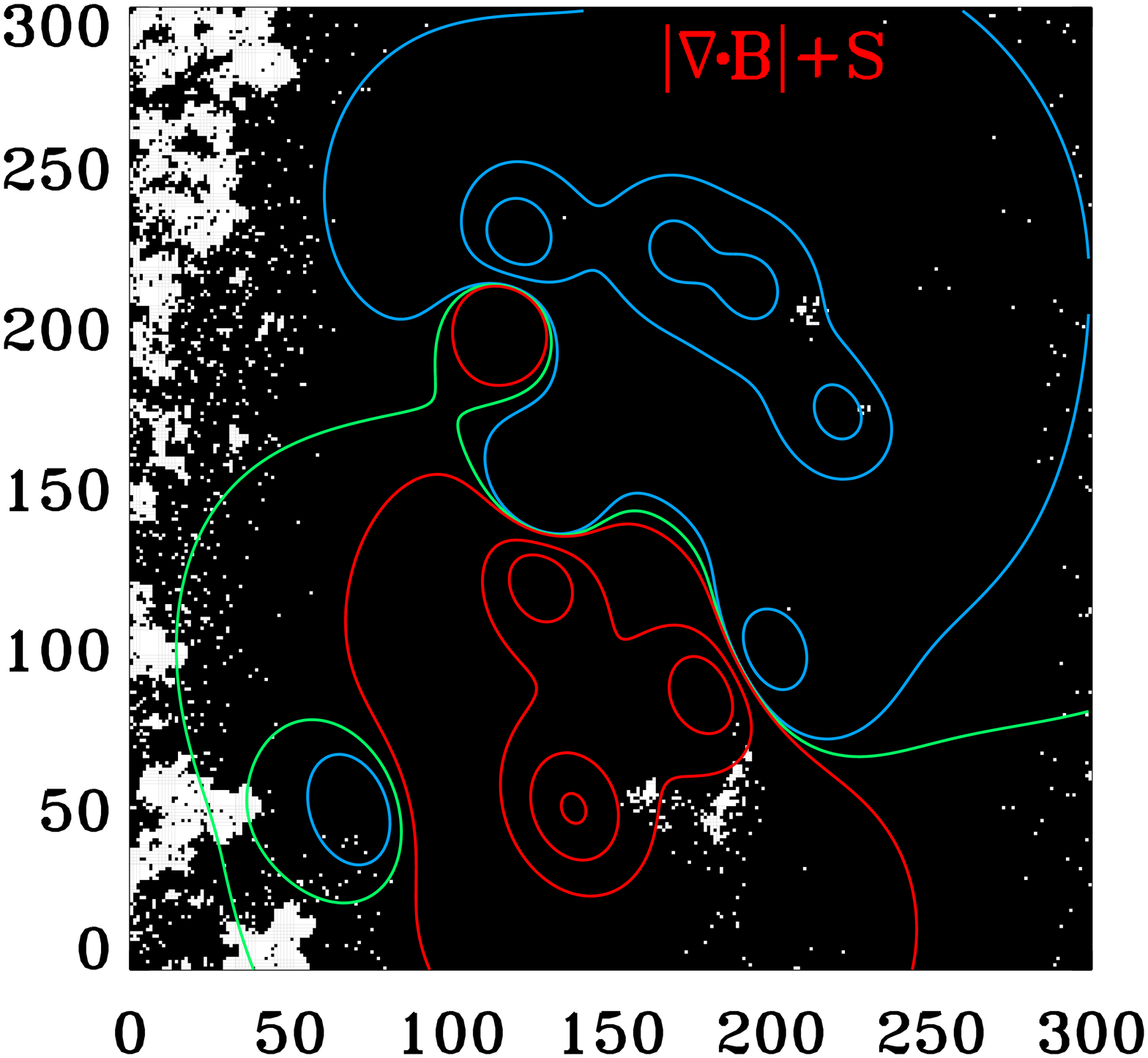} \\
\includegraphics[width=0.3\textwidth]{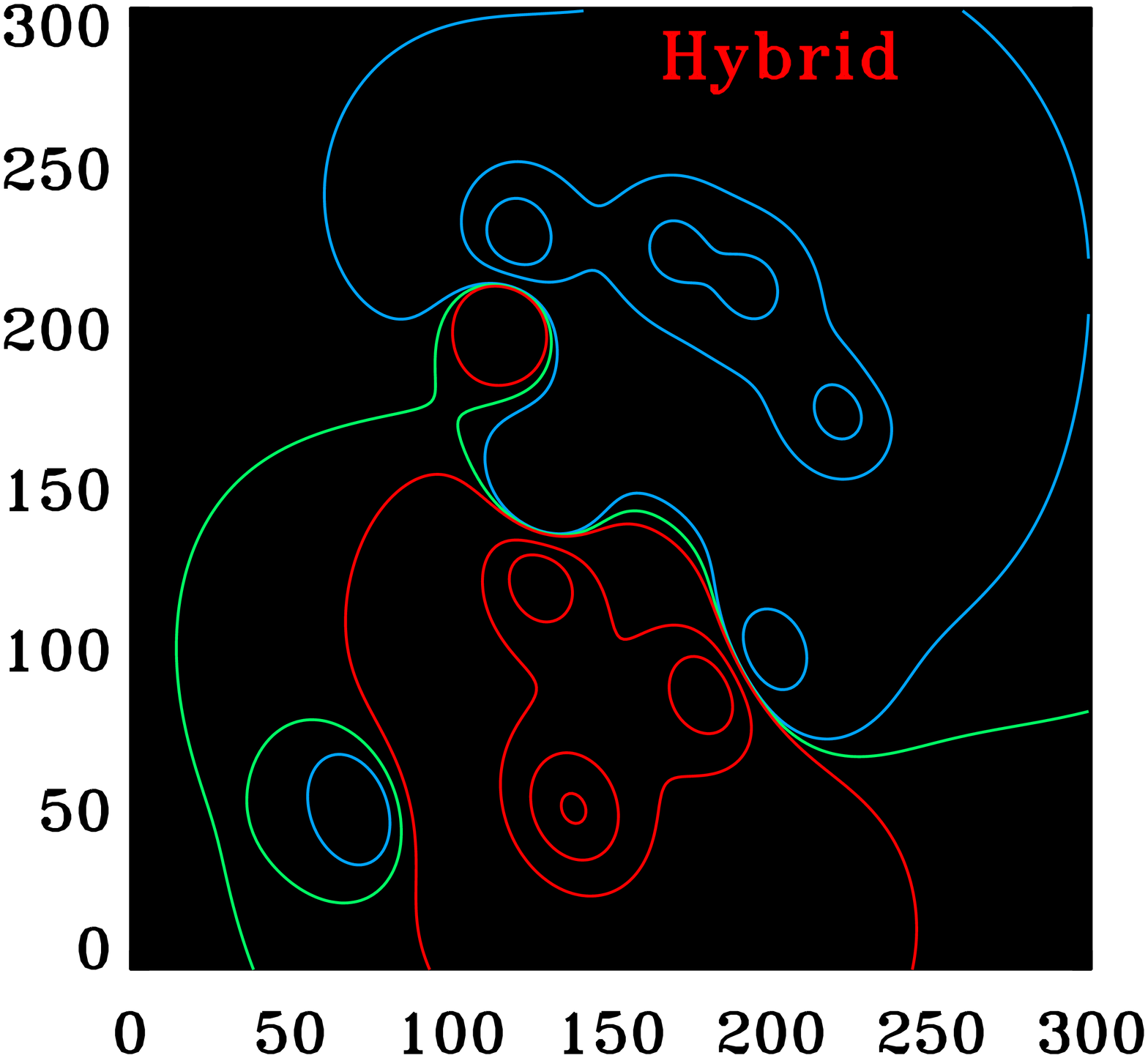} &
\includegraphics[width=0.3\textwidth]{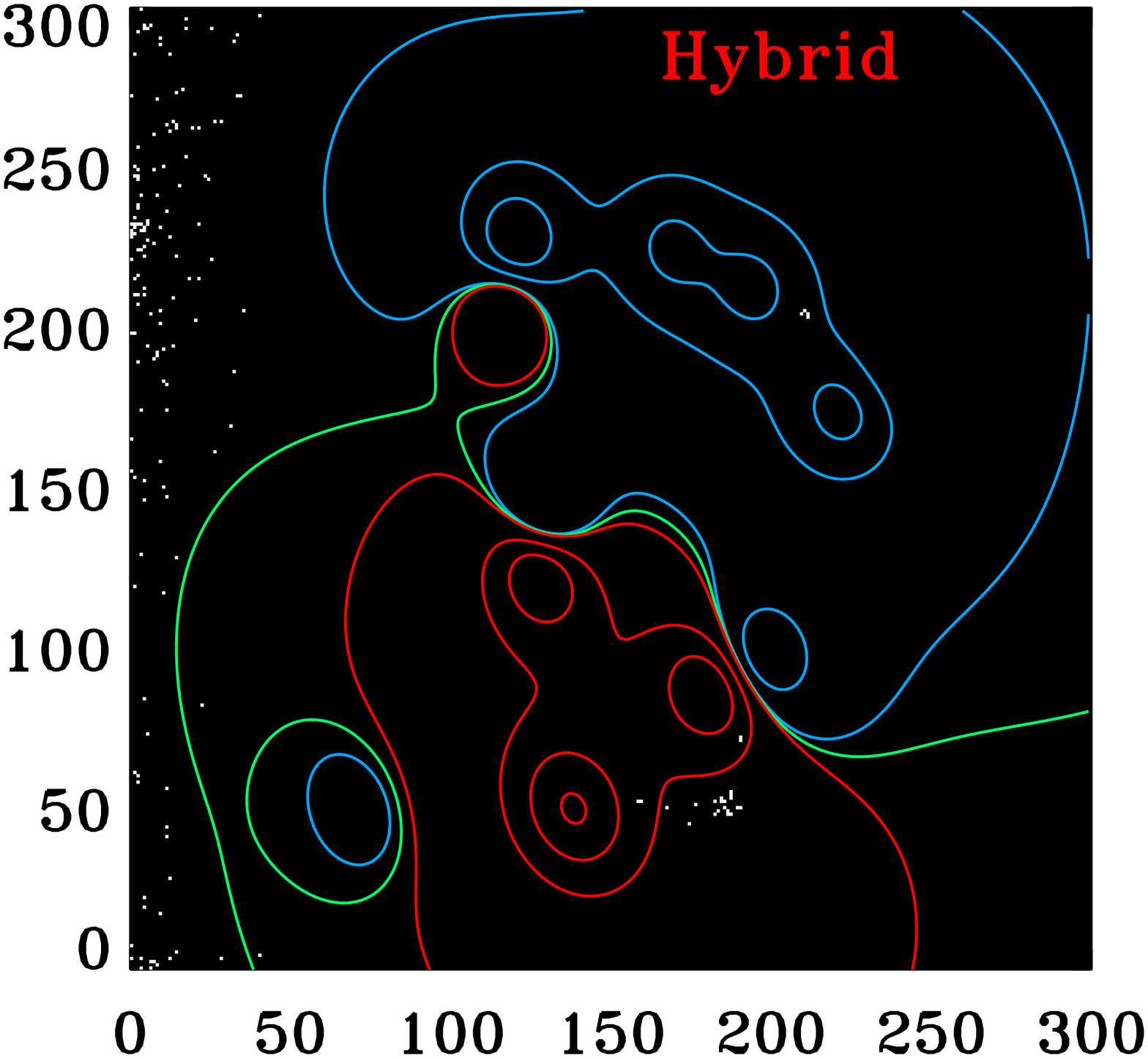} &
\includegraphics[width=0.3\textwidth]{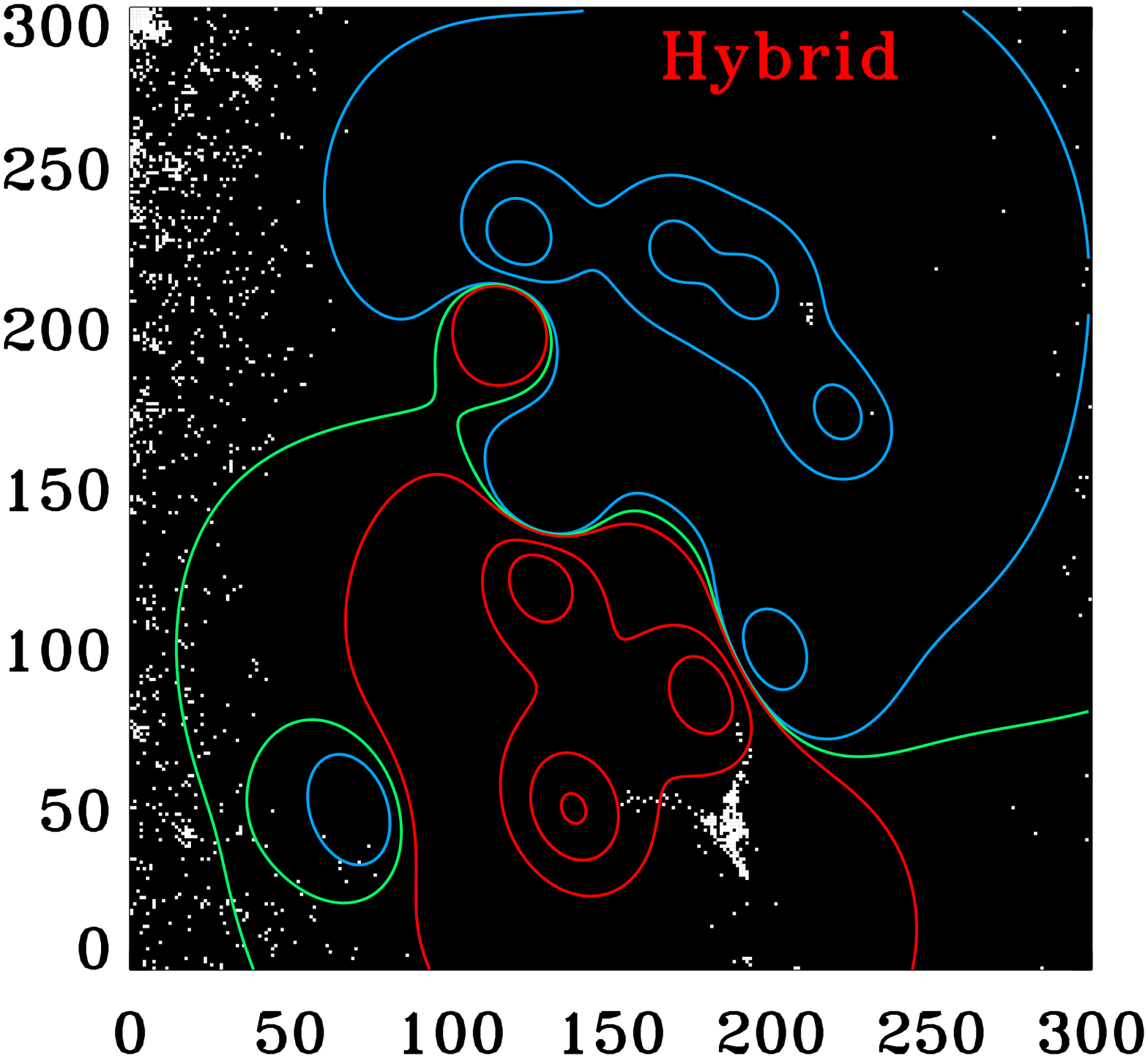}
\end{tabular}
\end{center}
\caption{Results for the various methods applied to the multipole field with different degrees of noise.
Areas with the correct/incorrect ambiguity resolution are black/white.
Positive/negative vertical magnetic flux for the noise-free case is indicated by red/blue contours at 50, 700, 1500 and 2800~G. 
The magnetic neutral line is indicated by the green contour.
(Left to right) No-noise, low-noise and high-noise cases.
(Top row) Results from the Wu and Ai (1990) criterion.
(Middle row) Results from the global minimisation method, which minimises \(E_{\rm s}\) (see Equation~(\ref{eds3})).
(Bottom row) Results from the hybrid method, which first minimises \(E_{\rm s}\) as in the middle row (see Equation~(\ref{eds3})), then applies a smoothing algorithm to pixels below a threshold transverse field strength (as described in Section~\ref{sec_hybrid}).
}
\label{tpd_fig}
\end{figure}

For the flux tube and arcade (Figure~\ref{fig_fan56}(a) and Table~\ref{tab_fan56}), the \inlinecite{1990AcApS..10..371W} criterion implemented with Equation~(\ref{wuaieqn}) performs significantly better than the implementation used in \cbl{} (with Equations~(\ref{divb2}) and (\ref{da})) in that the assumption made is correct over for most of the field of view.
It should be noted that the assumption made by the \inlinecite{1990AcApS..10..371W} criterion implemented with Equation~(\ref{wuaieqn})  is correct/incorrect when the correct/incorrect realisation of the azimuthal angle is selected; that is, the sign of the quantity \( \hs \sin \xi + \hc \cos \xi \) is correct for the correct choice of azimuthal angle.
Likewise, for the multipole field without noise (Figure~\ref{tpd_fig} and Table~\ref{tpd_tab}), the \inlinecite{1990AcApS..10..371W} criterion implemented with Equation~(\ref{wuaieqn}) performs significantly better than the implementation that uses Equations~(\ref{divb2}) and (\ref{da}), as demonstrated in \cbl{} and \adc{}.
However, as the level of noise increases, the results produced by the \inlinecite{1990AcApS..10..371W} criterion implemented with Equation~(\ref{wuaieqn}) degrade dramatically (Figure~\ref{tpd_fig} and Table~\ref{tpd_tab}).
Undesirable results are also produced for the cases with limited spatial resolution (Figure~\ref{flowers_fig} and Table~\ref{flowers_tab}).

\begin{table}[ht]
\caption{Performance metrics for the various methods applied to the twisted flux tube and arcade.}
\label{tab_fan56}
\begin{tabular}{lccc}
\hline
 & \( \mathcal{M}_{\rm area} \) & \( \mathcal{M}_{B_\perp  > 100~\rm{G}} \) & \( \mathcal{M}_{B_\perp  > 500~\rm{G}} \) \\
\hline
\inlinecite{1990AcApS..10..371W}                    & 0.98 & 0.98 & 0.97 \\
Global minimisation, \( | \grad \vdot \B | + \st \) & 1.00 & 1.00 & 1.00 \\
(Equation~(\ref{eds3})) & & & \\
Hybrid                                              & 1.00 & 1.00 & 1.00 \\
\hline
\end{tabular}
\end{table}

We have investigated an approach that assumes that the divergence of the magnetic field is minimised at each pixel, approximating derivatives as described above.
This approach is non-iterative, non-sequential, and the solution at one location does not depend on the choice of azimuthal angles in neighbouring locations.
However, when implemented with Equations~(\ref{divbafd})~--~(\ref{hceqn}), it can be shown that such an approach produces identical results to the \inlinecite{1990AcApS..10..371W} criterion implemented with Equation~(\ref{wuaieqn}).

\section{The Hybrid Method}
\label{sec_hybrid}

In the previous section, we found that methods for which the solution at one location does not depend on the choice of azimuthal angles in neighbouring locations tend to produce undesirable results when either noise or unresolved structure are present in the magnetogram data.
In this section we adapt the two-step, hybrid method that was presented in \adc{}, for which the solution at one location does depend on the choice of azimuthal angles in neighbouring locations.
The first step of this approach involves the global minimisation of a combination of the approximation for the divergence and a smoothness constraint.
In the second step, pixels with a weaker transverse component of the magnetic field are revisited with a smoothing algorithm.

\subsection{Step One: The Global Minimisation Method}

For the first step of the hybrid method, we assume that the correct configuration of azimuthal angles corresponds to the minimum of 

\begin{equation}
E_{\rm s} = \sum_{i=1}^{n_x}  \sum_{j=1}^{n_y} \left( | ( \grad \vdot \B )_{i,j,k} | + \ls \st_{i,j,k} \right) \, ,
\label{eds3}
\end{equation}

\noindent
where \( ( \grad \vdot \B )_{i,j,k} \) is the approximation for the  divergence of the magnetic field at pixel $(i, j, k)$, implemented using Equations~(\ref{divbafd})~--~(\ref{hceqn}) with derivatives approximated as described in Section~\ref{sec_wuai},  \(\ls\) is a dimensionless parameter, and \(\st_{i,j,k}\) is given by

\begin{eqnarray}
\st_{i,j,k} & = & \frac{1}{d_1} \left\{ \left[ \bxi \left( i, j, k \right) - \bxi \left( i+1, j, k \right) \right]^2 + \left[ \byi \left( i, j, k \right) - \byi \left( i+1, j, k \right) \right]^2 \right\}^{1/2} \nonumber\\
& + & \frac{1}{d_2}  \left\{ \left[ \bxi \left( i, j, k \right) - \bxi \left( i, j+1, k \right) \right]^2 + \left[ \byi \left( i, j, k \right) - \byi \left( i, j+1, k \right) \right]^2 \right\}^{1/2} \, , \label{sijk}
\end{eqnarray}

\noindent
where \(d_1\) is the distance between the centres of the pixels at \( \left( i, j, k \right) \) and \( \left(  i+1, j, k \right) \), and 
\(d_2\) is the distance between the centres of the pixels at \( \left( i, j, k \right) \) and \( \left(  i, j+1, k \right) \).
At the boundaries of the field of view, where $i=n_x$ and $j=n_y$, the definition of \( \st_{i,j,k}\) is modified to use only pixels within the field of view.
The purpose of \(\st_{i,j,k}\) in Equation~(\ref{eds3}) is to minimise the difference between the magnetic field in neighbouring pixels.
Thus, at a given pixel the first term in Equation~(\ref{eds3}) does not depend on choice of azimuthal angles in neighbouring pixels whereas the second term does (in the horizontal heliographic directions but not in the line-of-sight direction).
Because the definition of \(\st_{i,j,k}\) (see Equation~\ref{sijk}) involves only measurements of the magnetic field at a single height \(k\), each height can be disambiguated independently.
Consequently, this approach is less computationally expensive than the corresponding approach presented in \adc{}, which disambiguates all heights simultaneously because it was based on Equations~(\ref{divb2}) and (\ref{da}).
For example, to disambiguate two heights using this approach takes about half the compute time of the corresponding approach presented in \adc{}.

\begin{table}
\caption{Performance metrics for the various methods applied to the multipole field with different degrees of noise.}
\label{tpd_tab}
\begin{tabular}{llll|lll|lll}
\hline
Metric:
& \multicolumn{3}{c}{\( \mathcal{M}_{\rm area} \)}
& \multicolumn{3}{c}{\( \mathcal{M}_{B_\perp  > 100~\rm{G}} \)}
& \multicolumn{3}{c}{\( \mathcal{M}_{B_\perp  > 500~\rm{G}} \)}
\\
Noise level: & None & Low & High & None & Low & High & None & Low & High \\ 
\hline
\inlinecite{1990AcApS..10..371W}                           & 0.92  & 0.61 & 0.54 & 0.95 & 0.71 & 0.57 & 0.97 & 0.80 & 0.61 \\ 
Global minimisation, \( | \grad \vdot \B | + \st \)         & 1.00 & 0.99 & 0.94 & 1.00 & 1.00 & 0.98 & 1.00 & 1.00 & 1.00 \\ 
(Equation~(\ref{eds3})) & & &  & & & & & & \\
Hybrid                                                      & 1.00 & 1.00 & 0.99 & 1.00 & 1.00 & 1.00 & 1.00 & 1.00 & 1.00 \\
\hline
\end{tabular}
\end{table}

\begin{figure}[ht]
\begin{center}
\begin{tabular}{c@{\hspace{0.005\textwidth}}c@{\hspace{0.005\textwidth}}c}
\includegraphics[width=0.3\textwidth]{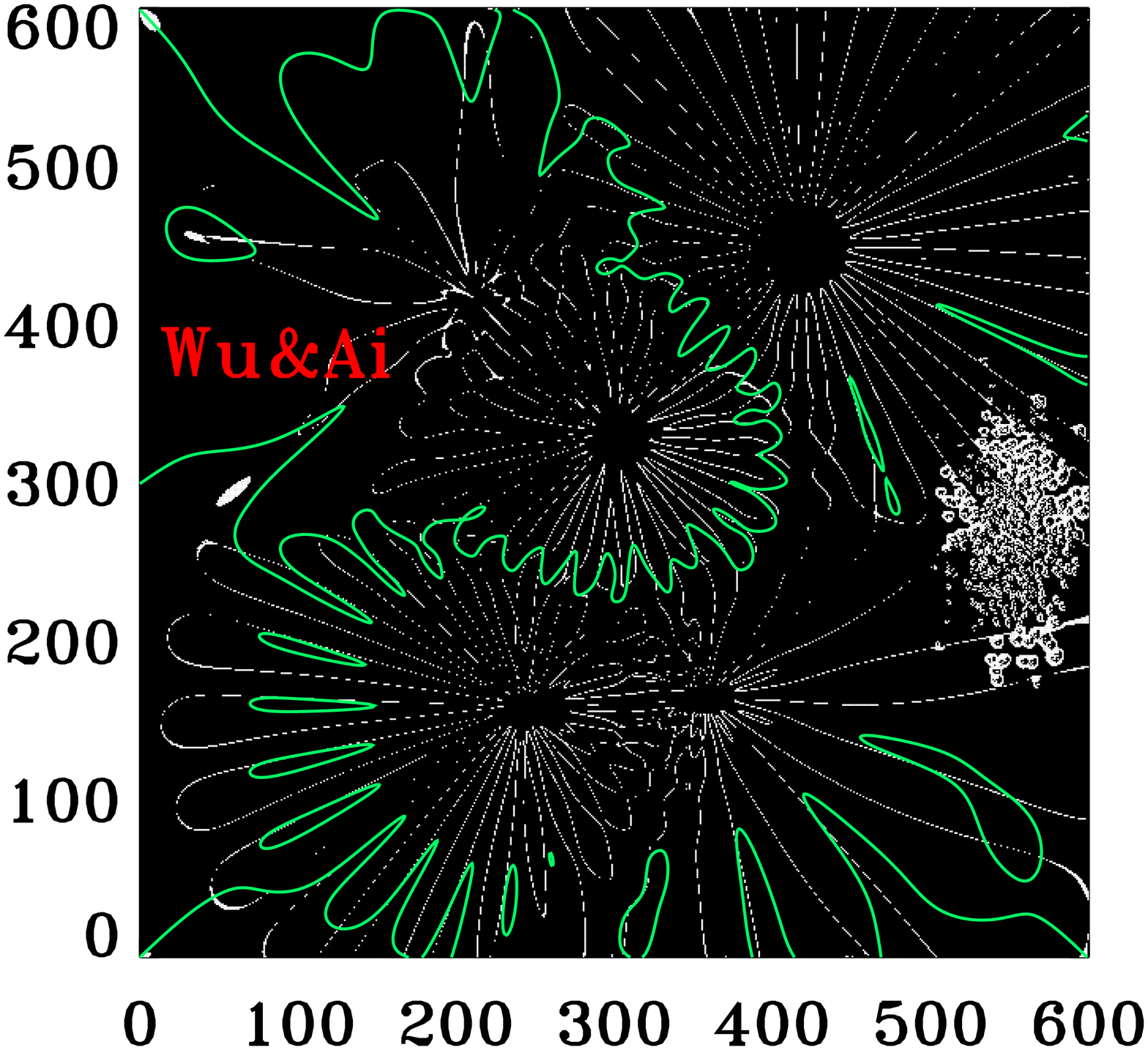} &
\includegraphics[width=0.3\textwidth]{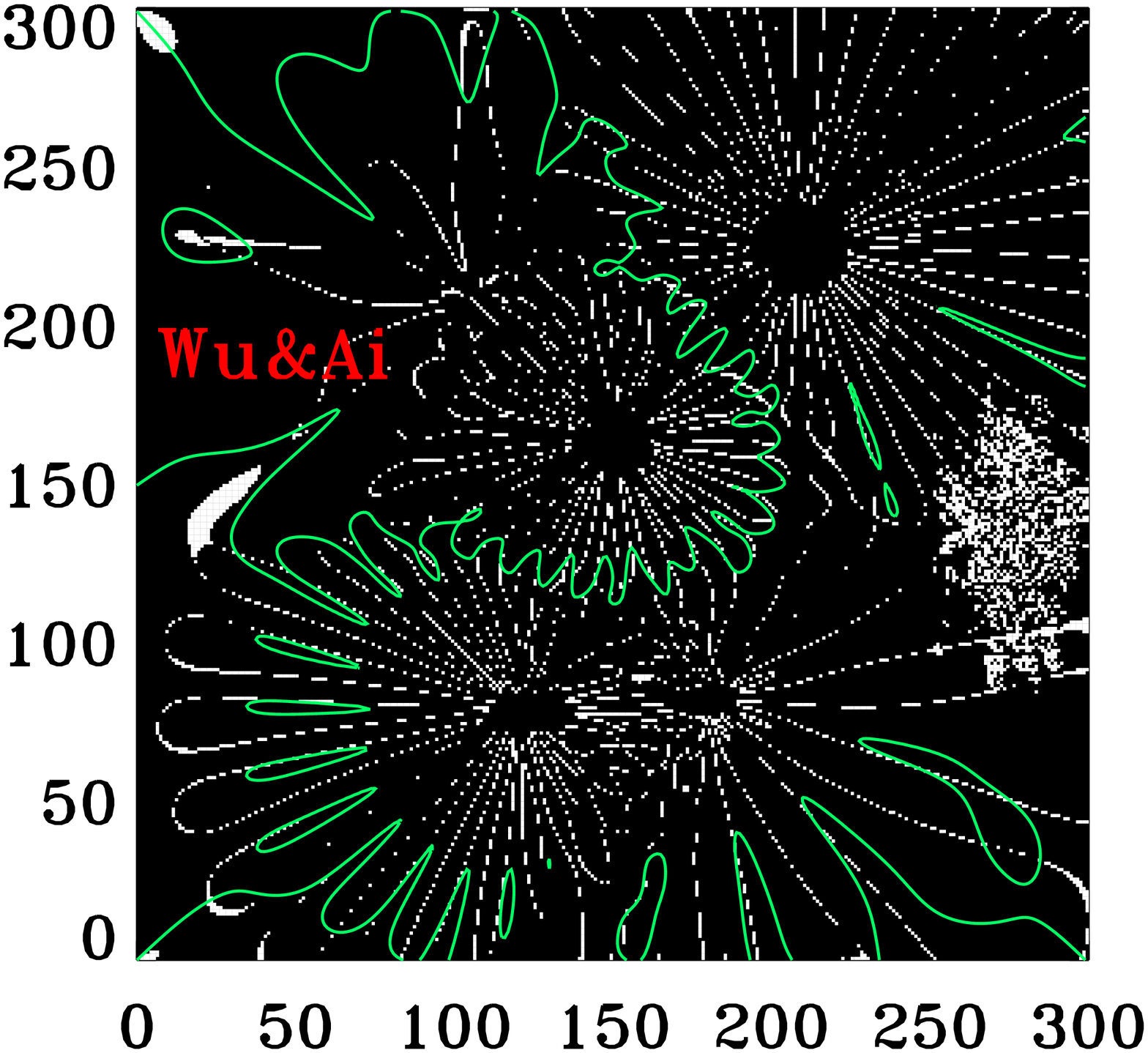} &
\includegraphics[width=0.3\textwidth]{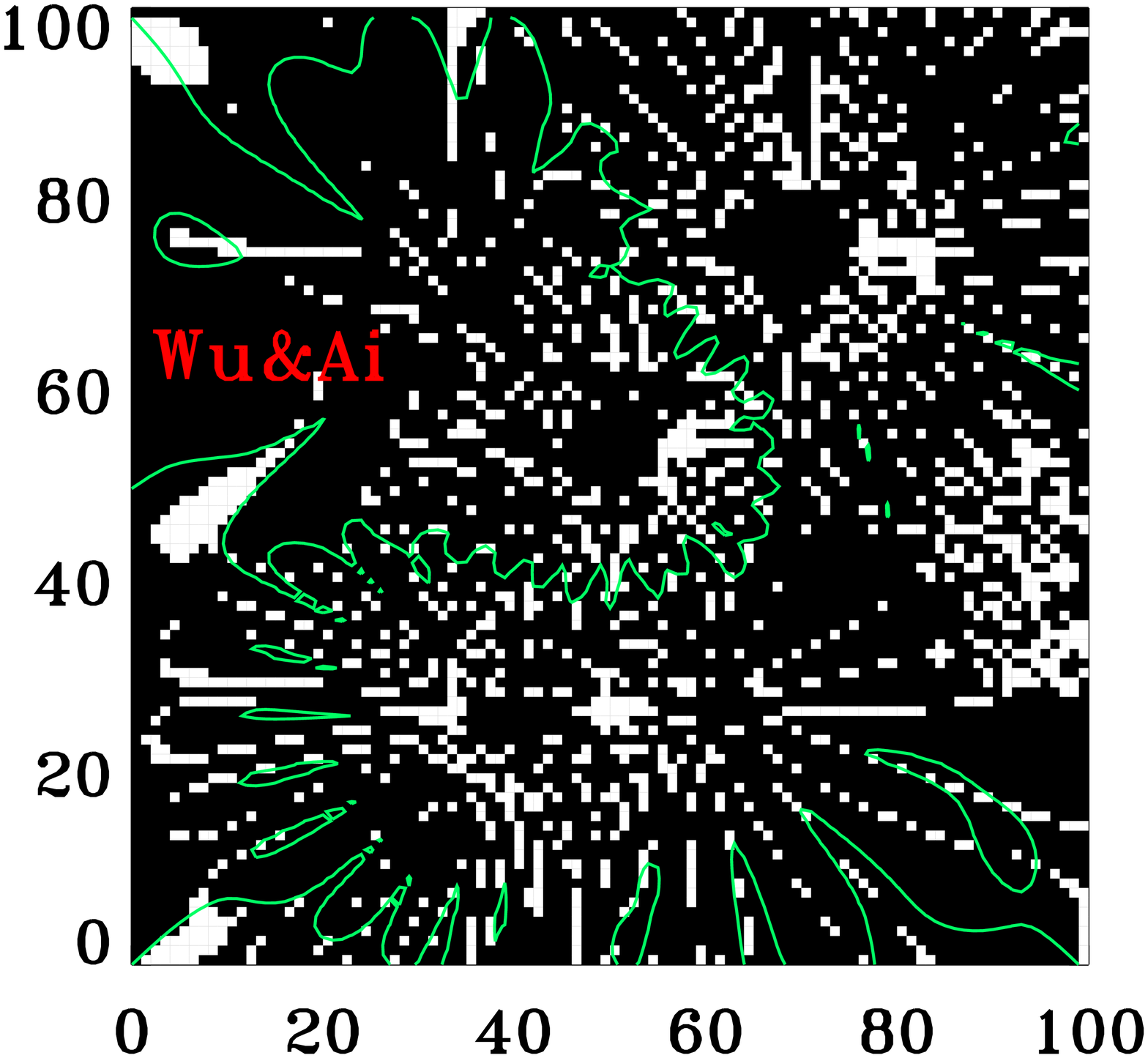} \\ 
\includegraphics[width=0.3\textwidth]{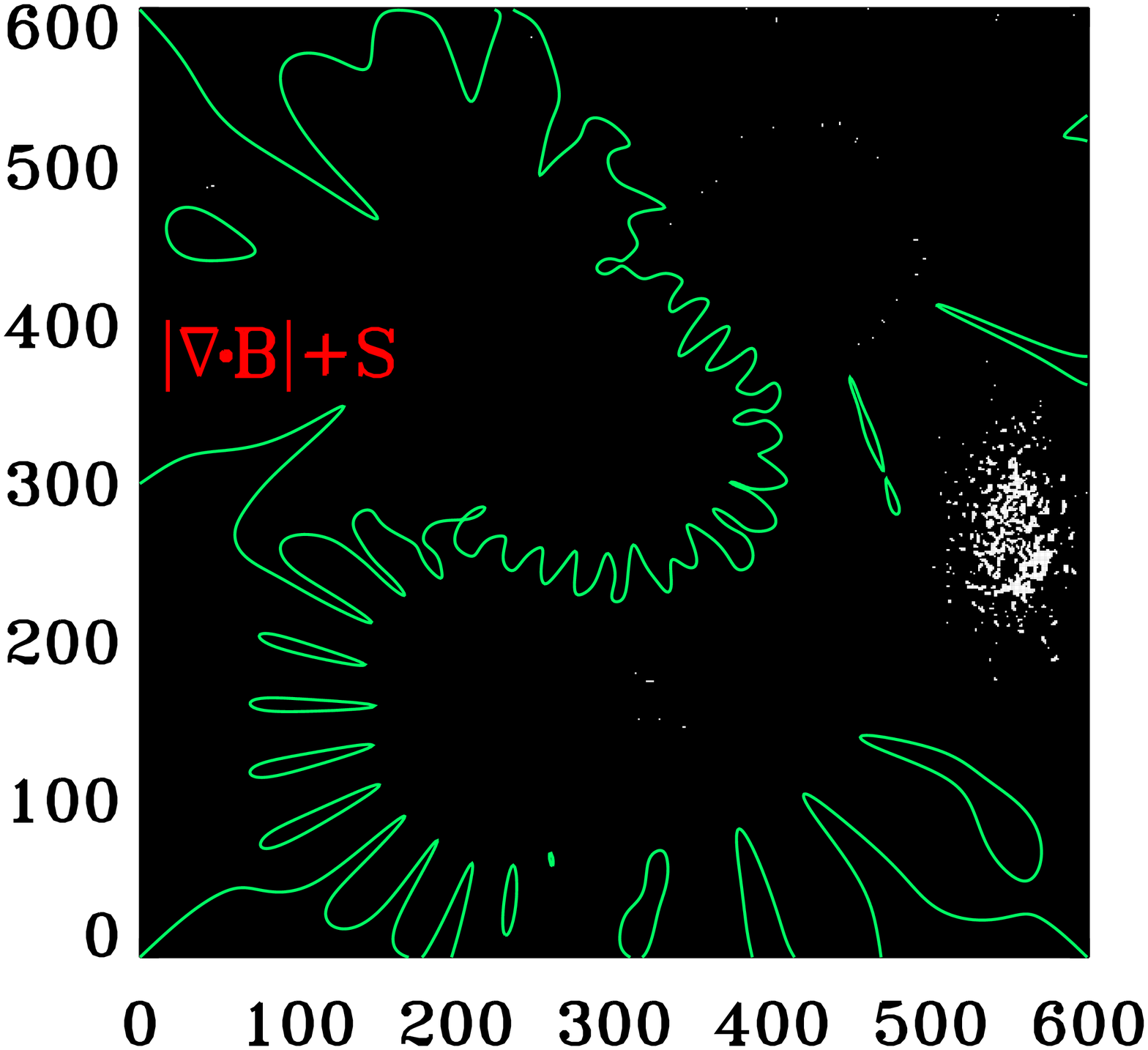} &
\includegraphics[width=0.3\textwidth]{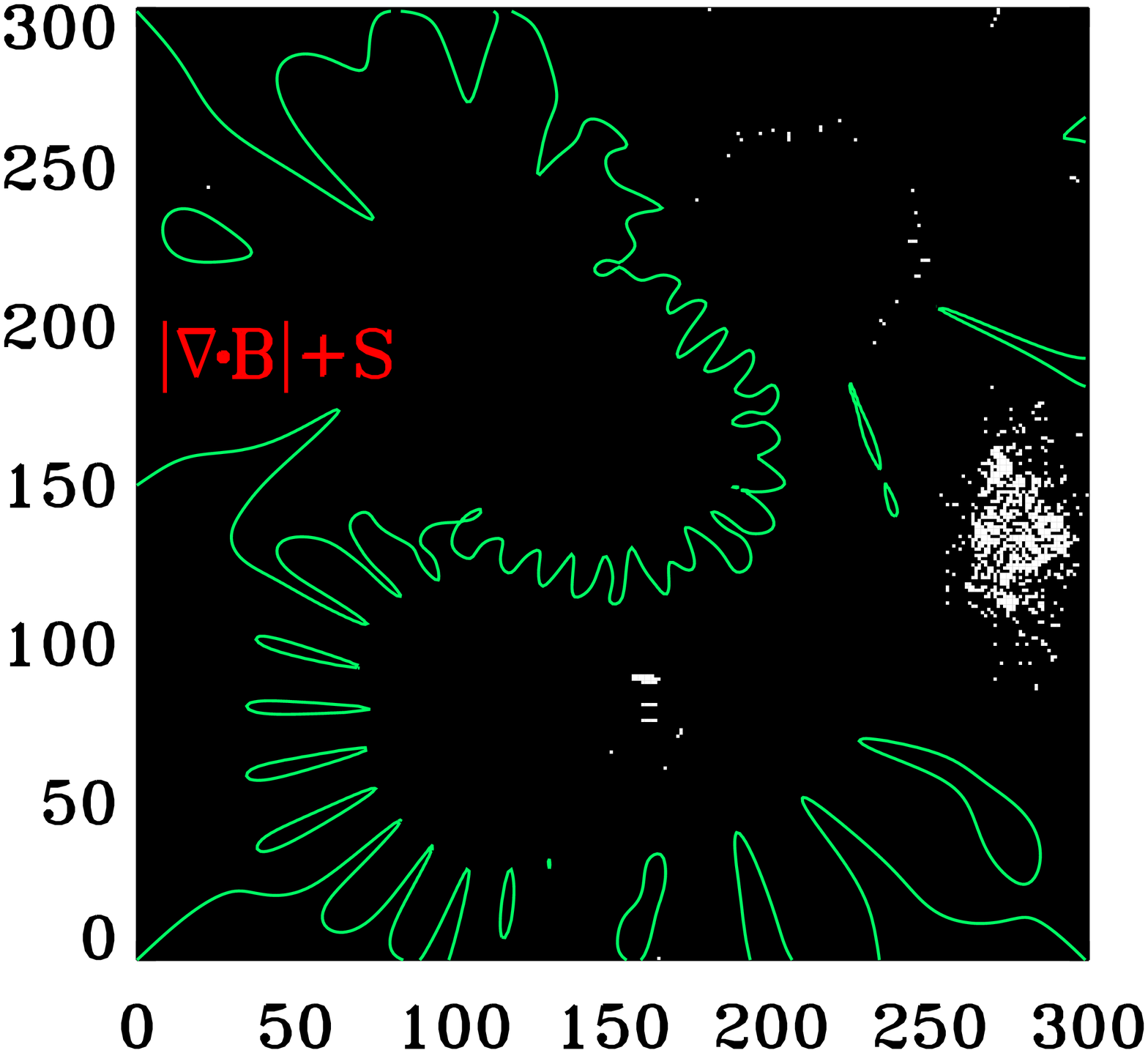} &
\includegraphics[width=0.3\textwidth]{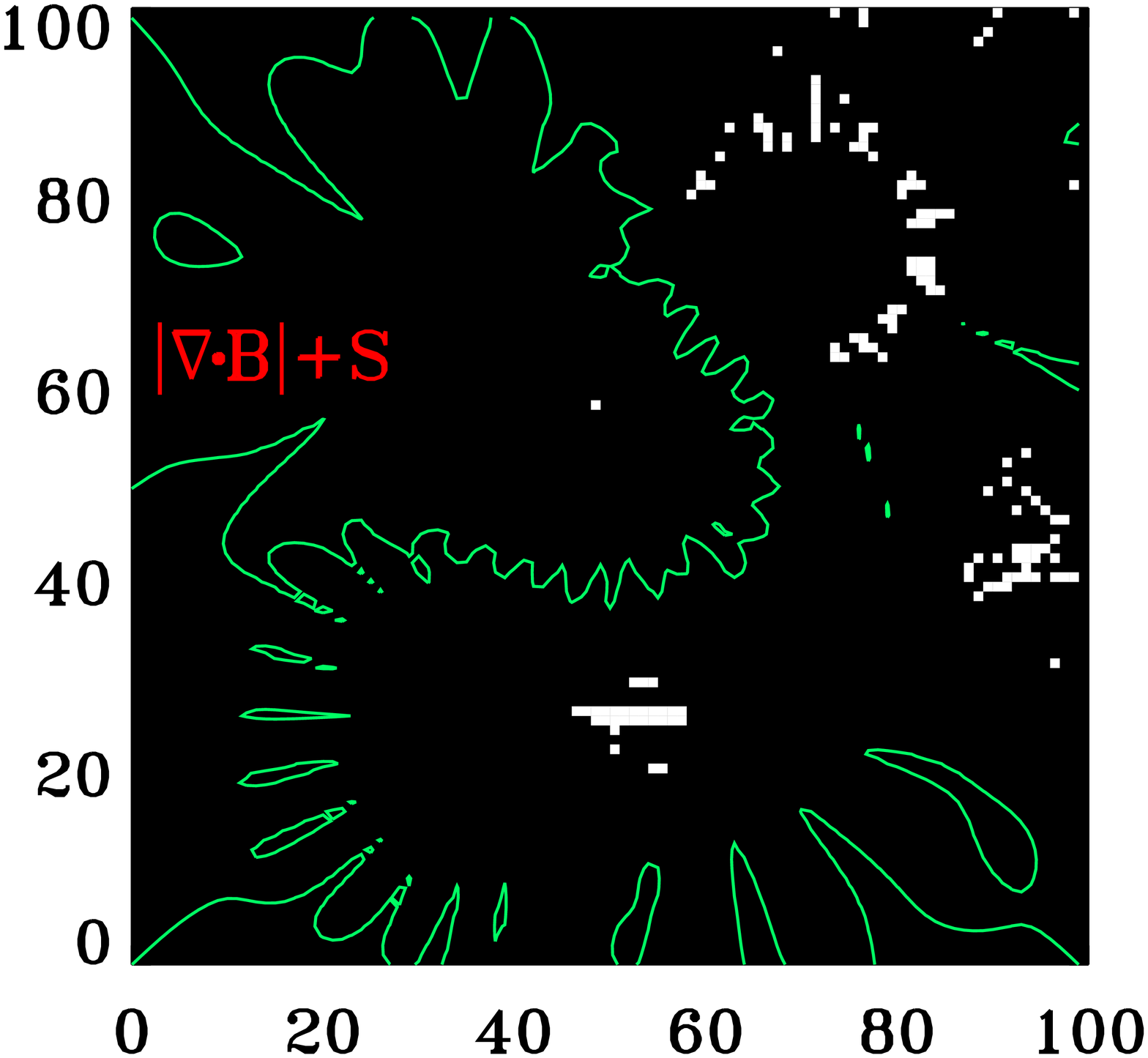} \\
\includegraphics[width=0.3\textwidth]{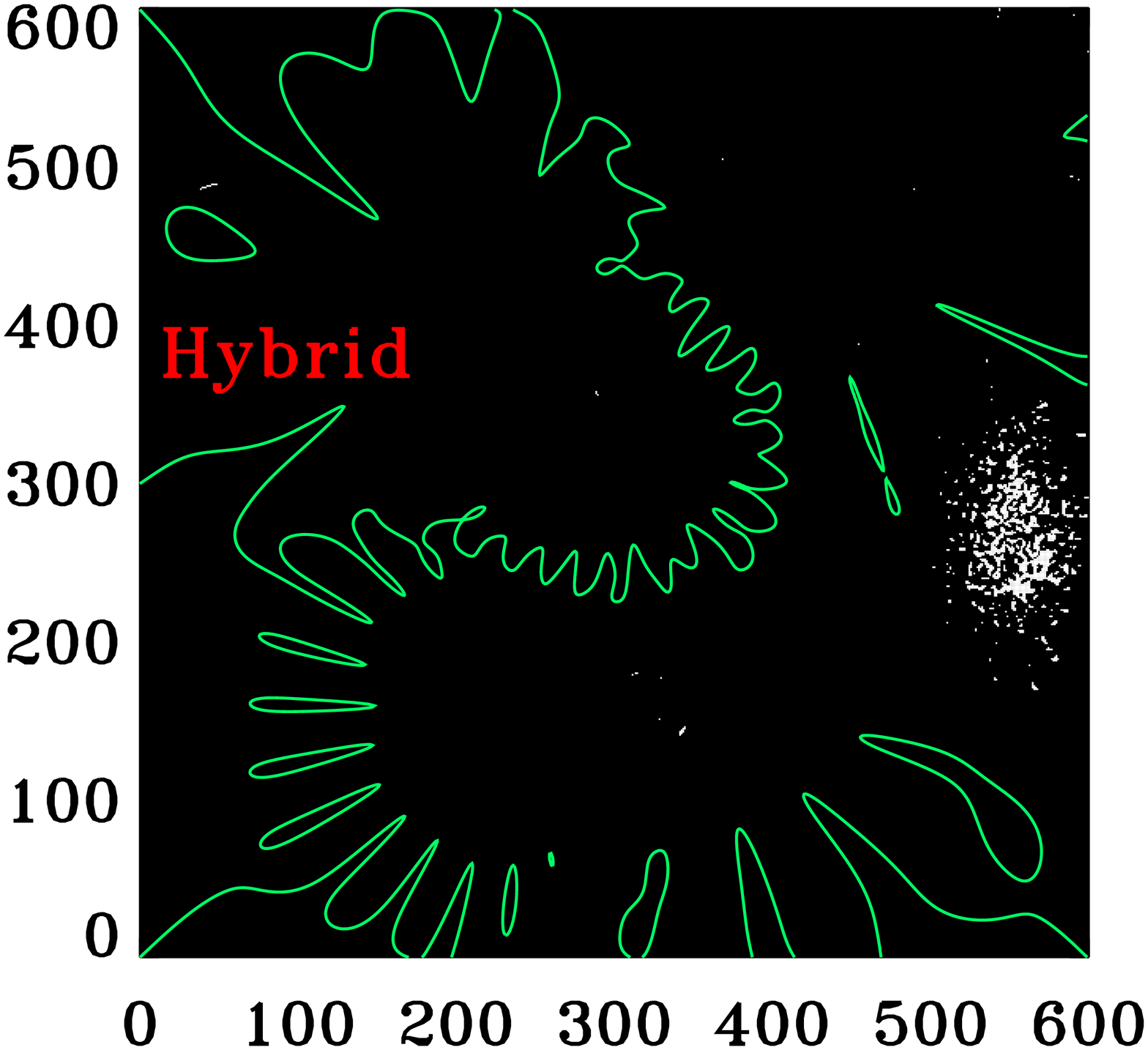} &
\includegraphics[width=0.3\textwidth]{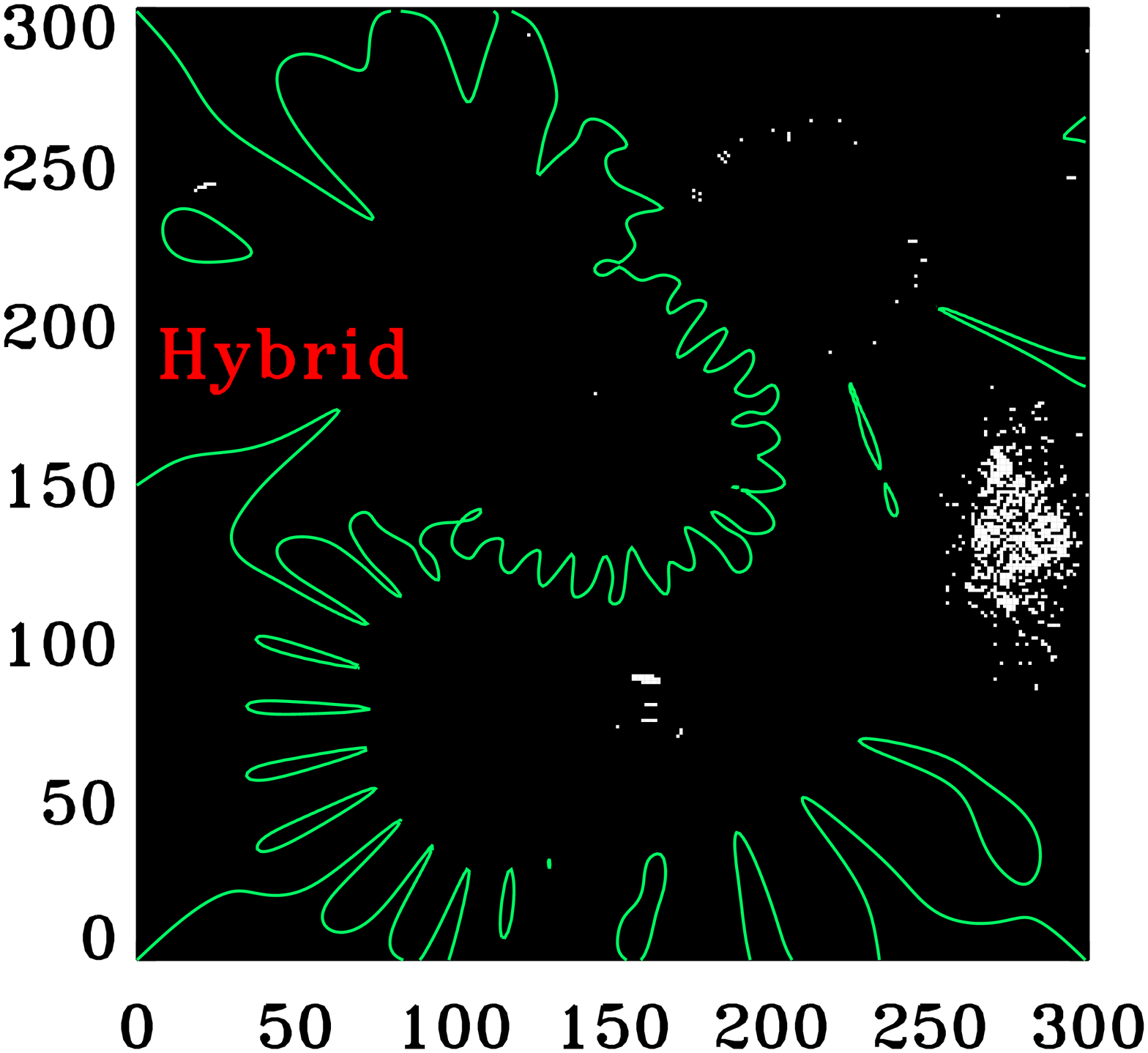} &
\includegraphics[width=0.3\textwidth]{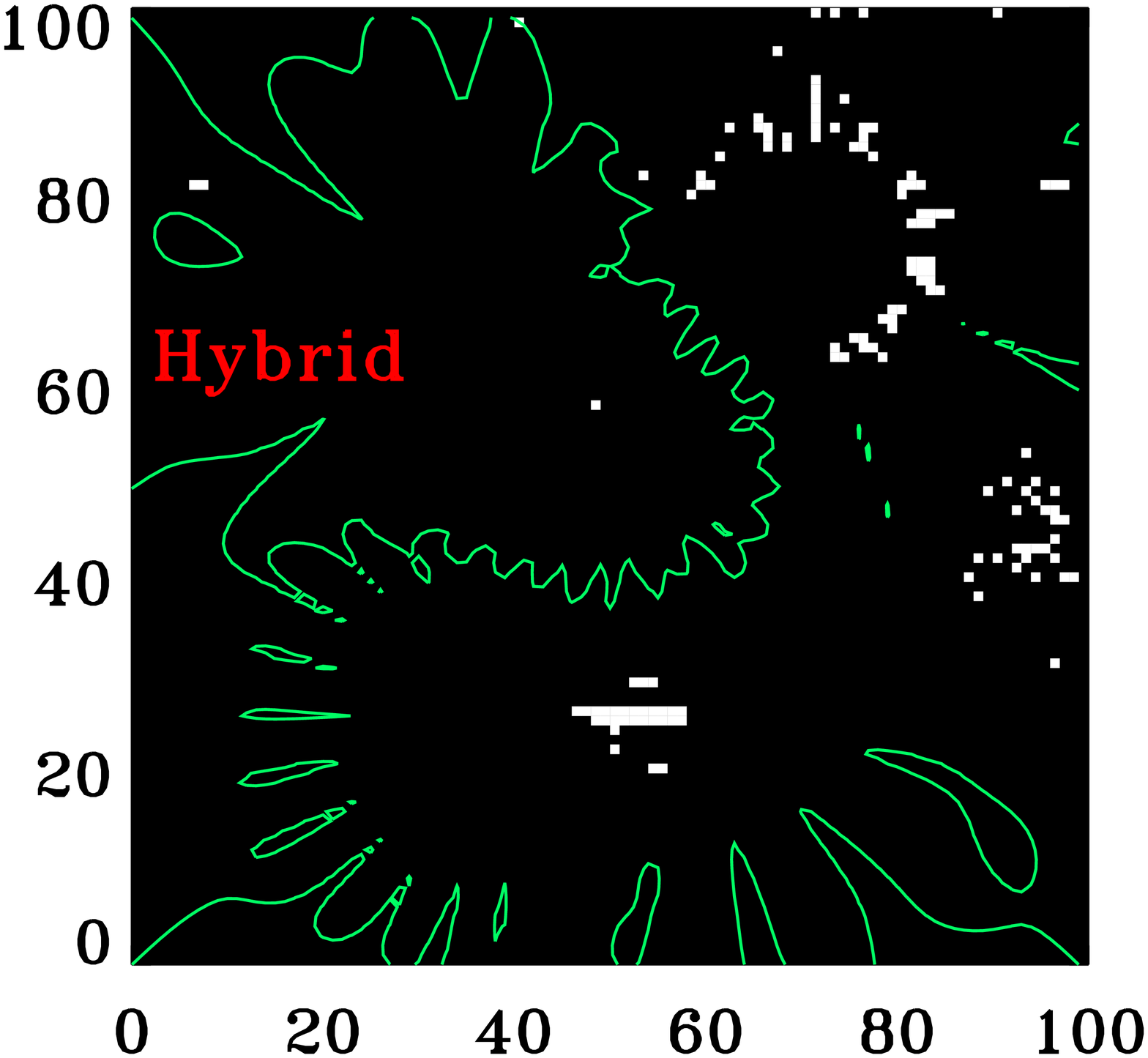}
\end{tabular}
\end{center}
\caption{Results for the various methods applied to the limited resolution cases.
Areas with the correct/incorrect ambiguity resolution are black/white.
The green contour is the magnetic neutral line.
(Left to right)  $0.15''$, $0.3''$ and $0.9''$ cases.
(Top row) Results from the Wu and Ai (1990) criterion.
(Middle row) Results from the global minimisation method, which minimises \(E_{\rm s}\) (see Equation~(\ref{eds3})).
(Bottom row) Results from the hybrid method, which first minimises \(E_{\rm s}\) as in the middle row (see Equation~(\ref{eds3})), then applies a smoothing algorithm to pixels below a threshold transverse field strength (as described in Section~\ref{sec_hybrid}).
}
\label{flowers_fig}
\end{figure}

To find the configuration of azimuthal angles that corresponds to the minimum of \(E_{\rm s}\) we use simulated annealing (\myeg \opencite{1953JChPh..21.1087M}; \opencite{1983Sci...220..671K}; \opencite{1992nrfa.book.....P}).
The simulated annealing algorithm is implemented exactly as described  in \adc{}.
We disambiguate each magnetogram twenty times with different sequences of random numbers and select the solution that yields the lowest value of \(E_{\rm s}\).
The value of the parameter \(\ls\) does affect the solution obtained; for simplicity, we set \(\ls=1\) for all cases.
Results from this first step of the hybrid method are presented in Figure~\ref{fig_fan56}(b), Figure~\ref{tpd_fig} (middle row), and Figure~\ref{flowers_fig} (middle row), and Tables~\ref{tab_fan56}, \ref{tpd_tab}, and \ref{flowers_tab}.

For the flux tube and arcade (Figure~\ref{fig_fan56}(b) and Table~\ref{tab_fan56}) many of the solutions retrieved by the global minimisation method have a value of \(E_{\rm s}\) less than that for the correct solution.
This indicates that the assumption that the correct  configuration of azimuthal angles corresponds to the global minimum of  \(E_{\rm s}\)  (Equation~(\ref{eds3})) is violated in these cases.
However, for the solution with the lowest value of \(E_{\rm s}\), the incorrect azimuthal angle is only obtained in a small number of pixels.

\begin{table}
\caption{Performance metrics for the various methods applied to the limited resolution cases.}
\label{flowers_tab}
\begin{tabular}{llll|lll|lll}
\hline
Metric:
& \multicolumn{3}{c}{\( \mathcal{M}_{\rm area} \)} 
& \multicolumn{3}{c}{\( \mathcal{M}_{B_\perp  > 100~\rm{G}} \)} 
& \multicolumn{3}{c}{\( \mathcal{M}_{B_\perp  > 500~\rm{G}} \)}
\\
Pixel size: &   $0.15''$ & $0.3''$ & $0.9''$ 
            &   $0.15''$ & $0.3''$ & $0.9''$ 
            &   $0.15''$ & $0.3''$ & $0.9''$ \\
\hline
\inlinecite{1990AcApS..10..371W}                                                              &     0.96 & 0.93 & 0.84 &   0.96 & 0.93 & 0.82  &   0.96 & 0.93 & 0.81 \\ 
Global minimisation, \( | \grad \vdot \B | + \st \)                                           &     0.99 & 0.99 & 0.99 &   1.00 & 1.00 & 0.99  &   1.00 & 1.00 & 1.00 \\ 
(Equation~(\ref{eds3})) & & &  & & & & & & \\
Hybrid                                                                                        &     0.99 & 0.99 & 0.99 &   1.00 & 1.00 & 0.99  &   1.00 & 1.00 & 1.00 \\
\hline
\end{tabular}
\end{table}

For the multipole field without noise (Figure~\ref{tpd_fig} and Table~\ref{tpd_tab}) the global minimisation method retrieves the correct solution at every pixel.
However, as the noise level increases, the performance of the  global minimisation method decreases. 
This is most evident in regions where the magnitude for the transverse component of the field is relatively weak and the signal tends to be dominated by noise (\myeg \(\xui \approx 0 \); \(\xui \approx 300 \); and at \(\xui \approx 170 \) and \(\yui \approx 50 \)).
For both noise-added cases, the solutions retrieved by the global minimisation method have a value of \(E_{\rm s}\) less than that for the correct solution.
Nevertheless, the quality of the solutions retrieved by the global minimisation method that minimises \(E_{\rm s}\) is very similar to that obtained by the corresponding method from \adc{}.

For all of the cases with limited spatial resolution (Figure~\ref{flowers_fig} and  Table~\ref{flowers_tab}) we again find that the assumption made by the global minimisation method is violated (\myie the solutions retrieved  have values of \(E_{\rm s}\) less than that for the correct solution).
However, the quality of the solutions retrieved by the global minimisation method is generally comparable to the corresponding method from \adc{}.
The approach implemented here tends to do better than the corresponding method from \adc{} in the plage region (at the centre, right of the field over field), and slightly worse around the flux concentration at the top, right of the field of view and between the two flux concentrations in the bottom, centre of the field of view.
All of these are regions with significant unresolved structure (see Figure~2 of \adc{}) and pose a challenge for all methods (see \opencite{2009SoPh..260...83L}).

\subsection{Step Two: A Smoothing Algorithm}

Evidently, the global minimisation method produces much better results than the \inlinecite{1990AcApS..10..371W} criterion.
However, it produces undesirable results in regions
where the transverse component of the magnetic field is strongly affected by photon noise.
Consequently, in the second step of the hybrid method we apply a smoothing algorithm that revisits pixels  in regions where the transverse component of the  field is weak. 
The smoothing algorithm does not use the divergence-free condition.
In contrast, it aims to minimise the difference between the magnetic field in a small neighbourhood of pixels.
The smoothing algorithm is identical to the method described in Section~5.4 of \adc{}, except that the definition of the neighbourhood is modified to include only pixels from a single height.
This allows each height to be disambiguated independently.
For the no-noise, low-noise, and high-noise cases we revisit pixels with a transverse field strength of 100~G, 150~G, and 400~G, as in \adc{}.

For the flux tube and arcade (Figure~\ref{fig_fan56}(c) and Table~\ref{tab_fan56}), the smoothing algorithm corrects the solutions in the small number of pixels that were incorrectly resolved by the global minimisation method.
The resulting solution has the correct azimuthal angle for all pixels.
For the multipole field without noise (Figure~\ref{tpd_fig} and Table~\ref{tpd_tab}), the smoothing algorithm does not change the solution in any pixels.
For the noise added cases the smoothing algorithm substantially improves the solutions in the regions that are revisited.
For the cases with limited spatial resolution (Figure~\ref{flowers_fig} and  Table~\ref{flowers_tab}), the smoothing algorithm changes the solution in a small number of pixels in each case, but does not substantially change the overall quality of the solutions.
Generally speaking, the final solutions produced by the two-step, hybrid method implemented here are very similar in quality to  those produced by both the corresponding approach from \adc{} and the better performing methods examined in \inlinecite{2009SoPh..260...83L}.
The smoothing algorithm works well for the synthetic datasets used in this article, which are quite limited in spatial extent; however, we note that it may not be the optimal method for use over extended areas far from regions where the solution is obtained from the global minimisation method.

\section{Conclusions}
\label{sec_conc}

We continue the investigation of how to use the divergence-free condition to resolve the azimuthal ambiguity in vector magnetogram data.
In previous work (\cbl{} and \adc{}), all of the methods examined used an expression for the divergence of the magnetic field that involves differentiation of quantities that depend on the choice of azimuthal angle.
As a result, all heights used to approximate line-of-sight derivatives should generally be disambiguated simultaneously.
In contrast, in this article we investigate a set of methods that use an expression for the divergence that involves differentiation of quantities that do not depend on the choice of azimuthal angle (\myeg{} \opencite{1996SoPh..164..291S}; \opencite{1998A+A...331..383S}; \opencite{2003JKAS...36S...7S}).
This results in an expression for the divergence that can be used to disambiguate each height independently.

We test two methods that each make different assumptions about how to use the divergence-free condition to resolve the ambiguity:
i) The \inlinecite{1990AcApS..10..371W} criterion, which expresses the equation for the divergence-free condition as an inequality (\myeg \opencite{1993A+A...278..279C}; \opencite{1993A+A...279..214L}; \opencite{2007ApJ...654..675L}; \opencite{2008SoPh..247...25C}; \cbl{}; \adc{}); and
ii) The two-step, hybrid method from \adc{}, adapted to disambiguate each height independently, which substantially improves its computational efficiency.
The first step of the hybrid method involves the  minimisation of a combination of the approximation for the divergence and a smoothness constraint.
The second step revisits pixels with a weaker transverse component of the magnetic field  with a smoothing algorithm.
To test these methods we use three different types of synthetic data:
i) error-free synthetic data;
ii) synthetic data with noise to simulate Poisson photon noise in the observed polarization spectra; and 
iii) synthetic data that includes a spatial binning to simulate the effects of limited instrumental spatial resolution in the directions perpendicular to the line-of-sight.
For error-free synthetic data the \inlinecite{1990AcApS..10..371W} criterion implemented in this article performs better than the implementation in \cbl{} and \adc{}.
However, we again find that the results produced by the \inlinecite{1990AcApS..10..371W} criterion are very sensitive to both noise and unresolved structure.
We also find again that the hybrid method outperforms the \inlinecite{1990AcApS..10..371W} criterion.
Furthermore, the quality of the results  retrieved by  the  hybrid method implemented in this article are generally very similar to those retrieved by the corresponding method in \adc{}, yet the compute time taken to produce the results is substantially reduced.
The software for the global minimisation method and the hybrid method is available at \url{http://www.cora.nwra.com/~ash/ambig1.tar.gz}.

\begin{acks}
The author thanks Takashi Sakurai for an insightful suggestion.
The author also thanks K.D. Leka, Yuhong Fan, and Graham Barnes for providing the synthetic datasets used in this investigation.
This work was supported by funding from NASA under contracts NNH05CC49C/NNH05CC75C and NNH09CE60C.
\end{acks}


\begin{thebibliography}{38}
\ifx\bisbn     \undefined \def\bisbn  #1{ISBN #1}\fi
\ifx\binits    \undefined \def\binits#1{#1}\fi
\ifx\bauthor   \undefined \def\bauthor#1{#1}\fi
\ifx\batitle   \undefined \def\batitle#1{#1}\fi
\ifx\bjtitle   \undefined \def\bjtitle#1{\textit{#1}}\fi
\ifx\bvolume   \undefined \def\bvolume#1{\textbf{#1}}\fi
\ifx\byear     \undefined \def\byear#1{#1}\fi
\ifx\bissue    \undefined \def\bissue#1{#1}\fi
\ifx\bfpage    \undefined \def\bfpage#1{#1}\fi
\ifx\blpage    \undefined \def\blpage #1{#1}\fi
\ifx\burl      \undefined \def\burl#1{\textsf{#1}}\fi
\ifx\href      \undefined \def\href#1#2{\textsf{#2}}\fi
\ifx\betal     \undefined \def\betal{\textit{et al.}}\fi
\ifx\bctitle   \undefined \def\bctitle#1{#1}\fi
\ifx\beditor   \undefined \def\beditor#1{#1}\fi
\ifx\bbtitle   \undefined \def\bbtitle#1{\textit{#1}}\fi
\ifx\bedition  \undefined \def\bedition#1{#1}\fi
\ifx\bseriesno \undefined \def\bseriesno#1{\textbf{#1}}\fi
\ifx\blocation \undefined \def\blocation#1{#1}\fi
\ifx\bsertitle \undefined \def\bsertitle#1{\textit{#1}}\fi
\ifx\bsnm      \undefined \def\bsnm#1{#1}\fi
\ifx\bsuffix   \undefined \def\bsuffix#1{#1}\fi
\ifx\bparticle \undefined \def\bparticle#1{#1}\fi
\ifx\barticle  \undefined \def\barticle#1{}\fi
\ifx\binstitute  \undefined \def\binstitute#1{#1}\fi
\ifx\bpublisher  \undefined \def\bpublisher#1{#1}\fi
\ifx\doiurl    \undefined
  \def\doiurl#1{\href{http://dx.doi.org/#1}{\textsf{DOI}}}\fi
\ifx\arxivurl  \undefined
  \def\arxivurl#1{\href{http://arxiv.org/abs/#1}{\textsf{arXiv}}}\fi
\ifx\adsurl    \undefined
  \def\adsurl#1{\href{http://adsabs.harvard.edu/abs/#1}{\textsf{ADS}}}\fi
\ifx\botherref \undefined \def\botherref#1{}\fi
\ifx\url       \undefined \def\url#1{\textsf{#1}}\fi
\ifx\bchapter  \undefined \def\bchapter#1{}\fi
\ifx\bbook     \undefined \def\bbook#1{}\fi
\ifx\bcomment  \undefined \def\bcomment#1{#1}\fi
\ifx\oauthor   \undefined \def\oauthor#1{#1}\fi
\ifx\citeauthoryear \undefined\def \citeauthoryear#1{#1}\fi
\def\endbibitem {}
\ifx\bconflocation  \undefined \def\bconflocation#1{#1} \fi

\bibitem[\protect\citeauthoryear{{Boulmezaoud} and
  {Amari}}{1999}]{1999A+A...347.1005B}
\begin{barticle}
\bauthor{\bsnm{{Boulmezaoud}}, \binits{T.Z.}},
\bauthor{\bsnm{{Amari}}, \binits{T.}}:
\byear{1999},
\bjtitle{\aap}
\bvolume{347},
\bfpage{1005}.
\adsurl{1999A\%26A...347.1005B}.
\end{barticle}
\endbibitem

\bibitem[\protect\citeauthoryear{{Collados}
  \textit{et~al.}}{1994}]{1994A+A...291..622C}
\begin{barticle}
\bauthor{\bsnm{{Collados}}, \binits{M.}},
\bauthor{\bsnm{{Martinez Pillet}}, \binits{V.}},
\bauthor{\bsnm{{Ruiz Cobo}}, \binits{B.}},
\bauthor{\bsnm{{del Toro Iniesta}}, \binits{J.C.}},
\bauthor{\bsnm{{Vazquez}}, \binits{M.}}:
\byear{1994},
\bjtitle{\aap}
\bvolume{291},
\bfpage{622}.
\adsurl{1994A\%26A...291..622C}.
\end{barticle}
\endbibitem

\bibitem[\protect\citeauthoryear{{Crouch}}{2013}]{2012SoPh..tmp..249C}
\begin{barticle}
\bauthor{\bsnm{{Crouch}}, \binits{A.D.}}:
\byear{2013},
\bjtitle{\solphys}
\bvolume{282},
\bfpage{107}.
\doiurl{10.1007/s11207-012-0149-8}.
\adsurl{2013SoPh..282..107C}.
\end{barticle}
\endbibitem

\bibitem[\protect\citeauthoryear{{Crouch} and
  {Barnes}}{2008}]{2008SoPh..247...25C}
\begin{barticle}
\bauthor{\bsnm{{Crouch}}, \binits{A.D.}},
\bauthor{\bsnm{{Barnes}}, \binits{G.}}:
\byear{2008},
\bjtitle{\solphys}
\bvolume{247},
\bfpage{25}.
\doiurl{10.1007/s11207-007-9096-1}.
\adsurl{2008SoPh..247...25C}.
\end{barticle}
\endbibitem

\bibitem[\protect\citeauthoryear{{Crouch}, {Barnes}, and
  {Leka}}{2009}]{2009SoPh..260..271C}
\begin{barticle}
\bauthor{\bsnm{{Crouch}}, \binits{A.D.}},
\bauthor{\bsnm{{Barnes}}, \binits{G.}},
\bauthor{\bsnm{{Leka}}, \binits{K.D.}}:
\byear{2009},
\bjtitle{\solphys}
\bvolume{260},
\bfpage{271}.
\doiurl{10.1007/s11207-009-9454-2}.
\adsurl{2009SoPh..260..271C}.
\end{barticle}
\endbibitem

\bibitem[\protect\citeauthoryear{{Cuperman}, {Li}, and
  {Semel}}{1993}]{1993A+A...278..279C}
\begin{barticle}
\bauthor{\bsnm{{Cuperman}}, \binits{S.}},
\bauthor{\bsnm{{Li}}, \binits{J.}},
\bauthor{\bsnm{{Semel}}, \binits{M.}}:
\byear{1993},
\bjtitle{\aap}
\bvolume{278},
\bfpage{279}.
\adsurl{1993A\%26A...278..279C}.
\end{barticle}
\endbibitem

\bibitem[\protect\citeauthoryear{{del Toro Iniesta} and {Ruiz
  Cobo}}{1996}]{1996SoPh..164..169D}
\begin{barticle}
\bauthor{\bsnm{{del Toro Iniesta}}, \binits{J.C.}},
\bauthor{\bsnm{{Ruiz Cobo}}, \binits{B.}}:
\byear{1996},
\bjtitle{\solphys}
\bvolume{164},
\bfpage{169}.
\doiurl{10.1007/BF00146631}.
\adsurl{1996SoPh..164..169D}.
\end{barticle}
\endbibitem

\bibitem[\protect\citeauthoryear{{Eibe}
  \textit{et~al.}}{2002}]{2002A+A...381..290E}
\begin{barticle}
\bauthor{\bsnm{{Eibe}}, \binits{M.T.}},
\bauthor{\bsnm{{Aulanier}}, \binits{G.}},
\bauthor{\bsnm{{Faurobert}}, \binits{M.}},
\bauthor{\bsnm{{Mein}}, \binits{P.}},
\bauthor{\bsnm{{Malherbe}}, \binits{J.M.}}:
\byear{2002},
\bjtitle{\aap}
\bvolume{381},
\bfpage{290}.
\doiurl{10.1051/0004-6361:20011495}.
\adsurl{2002A\%26A...381..290E}.
\end{barticle}
\endbibitem

\bibitem[\protect\citeauthoryear{{Fan} and {Gibson}}{2004}]{fan04}
\begin{barticle}
\bauthor{\bsnm{{Fan}}, \binits{Y.}},
\bauthor{\bsnm{{Gibson}}, \binits{S.E.}}:
\byear{2004},
\bjtitle{\apj}
\bvolume{609},
\bfpage{1123}.
\doiurl{10.1086/421238}.
\adsurl{2004ApJ...609.1123F}.
\end{barticle}
\endbibitem

\bibitem[\protect\citeauthoryear{{Gary} and
  {Hagyard}}{1990}]{1990SoPh..126...21G}
\begin{barticle}
\bauthor{\bsnm{{Gary}}, \binits{G.A.}},
\bauthor{\bsnm{{Hagyard}}, \binits{M.J.}}:
\byear{1990},
\bjtitle{\solphys}
\bvolume{126},
\bfpage{21}.
\adsurl{1990SoPh..126...21G}.
\end{barticle}
\endbibitem

\bibitem[\protect\citeauthoryear{{Georgoulis}}{2012}]{2012SoPh..276..423G}
\begin{barticle}
\bauthor{\bsnm{{Georgoulis}}, \binits{M.K.}}:
\byear{2012},
\bjtitle{\solphys}
\bvolume{276},
\bfpage{423}.
\doiurl{10.1007/s11207-011-9819-1}.
\adsurl{2012SoPh..276..423G}.
\end{barticle}
\endbibitem

\bibitem[\protect\citeauthoryear{{Harvey}}{1969}]{1969PhDT.........3H}
\begin{botherref}
\oauthor{\bsnm{{Harvey}}, \binits{J.W.}}:
1969,
{Magnetic Fields Associated with Solar Active-Region Prominences.}
Ph.D. thesis,
University of Colorado, Boulder.
\adsurl{1969PhDT.........3H}.
\end{botherref}
\endbibitem

\bibitem[\protect\citeauthoryear{{Kirkpatrick}, {Gelatt}, and
  {Vecchi}}{1983}]{1983Sci...220..671K}
\begin{barticle}
\bauthor{\bsnm{{Kirkpatrick}}, \binits{S.}},
\bauthor{\bsnm{{Gelatt}}, \binits{C.D.}},
\bauthor{\bsnm{{Vecchi}}, \binits{M.P.}}:
\byear{1983},
\bjtitle{Science}
\bvolume{220},
\bfpage{671}.
\doiurl{10.1126/science.220.4598.671}.
\adsurl{1983Sci...220..671K}.
\end{barticle}
\endbibitem

\bibitem[\protect\citeauthoryear{{Leka} and
  {Barnes}}{2012}]{2012SoPh..277...89L}
\begin{barticle}
\bauthor{\bsnm{{Leka}}, \binits{K.D.}},
\bauthor{\bsnm{{Barnes}}, \binits{G.}}:
\byear{2012},
\bjtitle{\solphys}
\bvolume{277},
\bfpage{89}.
\doiurl{10.1007/s11207-011-9821-7}.
\adsurl{2012SoPh..277...89L}.
\end{barticle}
\endbibitem

\bibitem[\protect\citeauthoryear{{Leka} and
  {Metcalf}}{2003}]{2003SoPh..212..361L}
\begin{barticle}
\bauthor{\bsnm{{Leka}}, \binits{K.D.}},
\bauthor{\bsnm{{Metcalf}}, \binits{T.R.}}:
\byear{2003},
\bjtitle{\solphys}
\bvolume{212},
\bfpage{361}.
\adsurl{2003SoPh..212..361L}.
\end{barticle}
\endbibitem

\bibitem[\protect\citeauthoryear{{Leka}, {Barnes}, and
  {Crouch}}{2009}]{2009ASPC..415..365L}
\begin{bchapter}
\bauthor{\bsnm{{Leka}}, \binits{K.D.}},
\bauthor{\bsnm{{Barnes}}, \binits{G.}},
\bauthor{\bsnm{{Crouch}}, \binits{A.}}:
\byear{2009},
In: \beditor{\bsnm{{Lites, B., Cheung, M., Magara, T., Mariska, J., Reeves,
  K.}}} (eds.)
\bbtitle{The Second Hinode Science Meeting: Beyond Discovery-Toward
  Understanding},
\bsertitle{ASP Conf. Ser.}
\bseriesno{415},
\bfpage{365}.
\adsurl{2009ASPC..415..365L}.
\end{bchapter}
\endbibitem

\bibitem[\protect\citeauthoryear{{Leka}
  \textit{et~al.}}{2009}]{2009SoPh..260...83L}
\begin{barticle}
\bauthor{\bsnm{{Leka}}, \binits{K.D.}},
\bauthor{\bsnm{{Barnes}}, \binits{G.}},
\bauthor{\bsnm{{Crouch}}, \binits{A.D.}},
\bauthor{\bsnm{{Metcalf}}, \binits{T.R.}},
\bauthor{\bsnm{{Gary}}, \binits{G.A.}},
\bauthor{\bsnm{{Jing}}, \binits{J.}},
\bauthor{\bsnm{{Liu}}, \binits{Y.}}:
\byear{2009},
\bjtitle{\solphys}
\bvolume{260},
\bfpage{83}.
\doiurl{10.1007/s11207-009-9440-8}.
\adsurl{2009SoPh..260...83L}.
\end{barticle}
\endbibitem

\bibitem[\protect\citeauthoryear{{Leka}
  \textit{et~al.}}{2012}]{2012SoPh..276..441L}
\begin{barticle}
\bauthor{\bsnm{{Leka}}, \binits{K.D.}},
\bauthor{\bsnm{{Barnes}}, \binits{G.}},
\bauthor{\bsnm{{Gary}}, \binits{G.A.}},
\bauthor{\bsnm{{Crouch}}, \binits{A.D.}},
\bauthor{\bsnm{{Liu}}, \binits{Y.}}:
\byear{2012},
\bjtitle{\solphys}
\bvolume{276},
\bfpage{441}.
\doiurl{10.1007/s11207-011-9879-2}.
\adsurl{2012SoPh..276..441L}.
\end{barticle}
\endbibitem

\bibitem[\protect\citeauthoryear{{Li}, {Amari}, and
  {Fan}}{2007}]{2007ApJ...654..675L}
\begin{barticle}
\bauthor{\bsnm{{Li}}, \binits{J.}},
\bauthor{\bsnm{{Amari}}, \binits{T.}},
\bauthor{\bsnm{{Fan}}, \binits{Y.}}:
\byear{2007},
\bjtitle{\apj}
\bvolume{654},
\bfpage{675}.
\doiurl{10.1086/509062}.
\adsurl{2007ApJ...654..675L}.
\end{barticle}
\endbibitem

\bibitem[\protect\citeauthoryear{{Li}, {Cuperman}, and
  {Semel}}{1993}]{1993A+A...279..214L}
\begin{barticle}
\bauthor{\bsnm{{Li}}, \binits{J.}},
\bauthor{\bsnm{{Cuperman}}, \binits{S.}},
\bauthor{\bsnm{{Semel}}, \binits{M.}}:
\byear{1993},
\bjtitle{\aap}
\bvolume{279},
\bfpage{214}.
\adsurl{1993A\%26A...279..214L}.
\end{barticle}
\endbibitem

\bibitem[\protect\citeauthoryear{{Liu}
  \textit{et~al.}}{1996}]{1996SoPh..169...79L}
\begin{barticle}
\bauthor{\bsnm{{Liu}}, \binits{Y.}},
\bauthor{\bsnm{{Wang}}, \binits{J.}},
\bauthor{\bsnm{{Yan}}, \binits{Y.}},
\bauthor{\bsnm{{Ai}}, \binits{G.}}:
\byear{1996},
\bjtitle{\solphys}
\bvolume{169},
\bfpage{79}.
\doiurl{10.1007/BF00153835}.
\adsurl{1996SoPh..169...79L}.
\end{barticle}
\endbibitem

\bibitem[\protect\citeauthoryear{{Maltby}
  \textit{et~al.}}{1986}]{1986ApJ...306..284M}
\begin{barticle}
\bauthor{\bsnm{{Maltby}}, \binits{P.}},
\bauthor{\bsnm{{Avrett}}, \binits{E.H.}},
\bauthor{\bsnm{{Carlsson}}, \binits{M.}},
\bauthor{\bsnm{{Kjeldseth-Moe}}, \binits{O.}},
\bauthor{\bsnm{{Kurucz}}, \binits{R.L.}},
\bauthor{\bsnm{{Loeser}}, \binits{R.}}:
\byear{1986},
\bjtitle{\apj}
\bvolume{306},
\bfpage{284}.
\doiurl{10.1086/164342}.
\adsurl{1986ApJ...306..284M}.
\end{barticle}
\endbibitem

\bibitem[\protect\citeauthoryear{{Metcalf}}{1994}]{1994SoPh..155..235M}
\begin{barticle}
\bauthor{\bsnm{{Metcalf}}, \binits{T.R.}}:
\byear{1994},
\bjtitle{\solphys}
\bvolume{155},
\bfpage{235}.
\doiurl{10.1007/BF00680593}.
\adsurl{1994SoPh..155..235M}.
\end{barticle}
\endbibitem

\bibitem[\protect\citeauthoryear{{Metcalf}
  \textit{et~al.}}{1995}]{1995ApJ...439..474M}
\begin{barticle}
\bauthor{\bsnm{{Metcalf}}, \binits{T.R.}},
\bauthor{\bsnm{{Jiao}}, \binits{L.}},
\bauthor{\bsnm{{McClymont}}, \binits{A.N.}},
\bauthor{\bsnm{{Canfield}}, \binits{R.C.}},
\bauthor{\bsnm{{Uitenbroek}}, \binits{H.}}:
\byear{1995},
\bjtitle{\apj}
\bvolume{439},
\bfpage{474}.
\doiurl{10.1086/175188}.
\adsurl{1995ApJ...439..474M}.
\end{barticle}
\endbibitem

\bibitem[\protect\citeauthoryear{{Metcalf}
  \textit{et~al.}}{2006}]{2006SoPh..237..267M}
\begin{barticle}
\bauthor{\bsnm{{Metcalf}}, \binits{T.R.}},
\bauthor{\bsnm{{Leka}}, \binits{K.D.}},
\bauthor{\bsnm{{Barnes}}, \binits{G.}},
\bauthor{\bsnm{{Lites}}, \binits{B.W.}},
\bauthor{\bsnm{{Georgoulis}}, \binits{M.K.}},
\bauthor{\bsnm{{Pevtsov}}, \binits{A.A.}},
\bauthor{\textit{et~al.}}:
\byear{2006},
\bjtitle{\solphys}
\bvolume{237},
\bfpage{267}.
\doiurl{10.1007/s11207-006-0170-x}.
\adsurl{2006SoPh..237..267M}.
\end{barticle}
\endbibitem

\bibitem[\protect\citeauthoryear{{Metropolis}
  \textit{et~al.}}{1953}]{1953JChPh..21.1087M}
\begin{barticle}
\bauthor{\bsnm{{Metropolis}}, \binits{N.}},
\bauthor{\bsnm{{Rosenbluth}}, \binits{A.W.}},
\bauthor{\bsnm{{Rosenbluth}}, \binits{M.N.}},
\bauthor{\bsnm{{Teller}}, \binits{A.H.}},
\bauthor{\bsnm{{Teller}}, \binits{E.}}:
\byear{1953},
\bjtitle{\jcp}
\bvolume{21},
\bfpage{1087}.
\doiurl{10.1063/1.1699114}.
\adsurl{1953JChPh..21.1087M}.
\end{barticle}
\endbibitem

\bibitem[\protect\citeauthoryear{{Press}
  \textit{et~al.}}{1992}]{1992nrfa.book.....P}
\begin{bbook}
\bauthor{\bsnm{{Press}}, \binits{W.H.}},
\bauthor{\bsnm{{Teukolsky}}, \binits{S.A.}},
\bauthor{\bsnm{{Vetterling}}, \binits{W.T.}},
\bauthor{\bsnm{{Flannery}}, \binits{B.P.}}:
\byear{1992},
\bbtitle{{Numerical Recipes in FORTRAN. The Art of Scientific Computing}},
\bpublisher{Cambridge University Press}, \blocation{Cambridge},
\bfpage{436},
\adsurl{1992nrfa.book.....P}.
\end{bbook}
\endbibitem

\bibitem[\protect\citeauthoryear{{Ruiz Cobo} and {del Toro
  Iniesta}}{1992}]{1992ApJ...398..375R}
\begin{barticle}
\bauthor{\bsnm{{Ruiz Cobo}}, \binits{B.}},
\bauthor{\bsnm{{del Toro Iniesta}}, \binits{J.C.}}:
\byear{1992},
\bjtitle{\apj}
\bvolume{398},
\bfpage{375}.
\doiurl{10.1086/171862}.
\adsurl{1992ApJ...398..375R}.
\end{barticle}
\endbibitem

\bibitem[\protect\citeauthoryear{{Sakurai} and
  {Hagino}}{2003}]{2003JKAS...36S...7S}
\begin{barticle}
\bauthor{\bsnm{{Sakurai}}, \binits{T.}},
\bauthor{\bsnm{{Hagino}}, \binits{M.}}:
\byear{2003},
\bjtitle{J. Korean Astron. Soc.}
\bvolume{36},
\bfpage{7}.
\adsurl{2003JKAS...36S...7S}.
\end{barticle}
\endbibitem

\bibitem[\protect\citeauthoryear{{Semel} and
  {Skumanich}}{1998}]{1998A+A...331..383S}
\begin{barticle}
\bauthor{\bsnm{{Semel}}, \binits{M.}},
\bauthor{\bsnm{{Skumanich}}, \binits{A.}}:
\byear{1998},
\bjtitle{\aap}
\bvolume{331},
\bfpage{383}.
\adsurl{1998A\%26A...331..383S}.
\end{barticle}
\endbibitem

\bibitem[\protect\citeauthoryear{{Skumanich} and
  {Semel}}{1996}]{1996SoPh..164..291S}
\begin{barticle}
\bauthor{\bsnm{{Skumanich}}, \binits{A.}},
\bauthor{\bsnm{{Semel}}, \binits{M.}}:
\byear{1996},
\bjtitle{\solphys}
\bvolume{164},
\bfpage{291}.
\doiurl{10.1007/BF00146641}.
\adsurl{1996SoPh..164..291S}.
\end{barticle}
\endbibitem

\bibitem[\protect\citeauthoryear{{Socas-Navarro}}{2005}]{2005ApJ...631L.167S}
\begin{barticle}
\bauthor{\bsnm{{Socas-Navarro}}, \binits{H.}}:
\byear{2005},
\bjtitle{\apjl}
\bvolume{631},
\bfpage{L167}.
\doiurl{10.1086/497334}.
\adsurl{2005ApJ...631L.167S}.
\end{barticle}
\endbibitem

\bibitem[\protect\citeauthoryear{{Socas-Navarro}}{2007}]{2007ApJS..169..439S}
\begin{barticle}
\bauthor{\bsnm{{Socas-Navarro}}, \binits{H.}}:
\byear{2007},
\bjtitle{\apjs}
\bvolume{169},
\bfpage{439}.
\doiurl{10.1086/510336}.
\adsurl{2007ApJS..169..439S}.
\end{barticle}
\endbibitem

\bibitem[\protect\citeauthoryear{{Socas-Navarro}, {Trujillo Bueno}, and {Ruiz
  Cobo}}{2000}]{2000ApJ...530..977S}
\begin{barticle}
\bauthor{\bsnm{{Socas-Navarro}}, \binits{H.}},
\bauthor{\bsnm{{Trujillo Bueno}}, \binits{J.}},
\bauthor{\bsnm{{Ruiz Cobo}}, \binits{B.}}:
\byear{2000},
\bjtitle{\apj}
\bvolume{530},
\bfpage{977}.
\doiurl{10.1086/308414}.
\adsurl{2000ApJ...530..977S}.
\end{barticle}
\endbibitem

\bibitem[\protect\citeauthoryear{{Vernazza}, {Avrett}, and
  {Loeser}}{1981}]{1981ApJS...45..635V}
\begin{barticle}
\bauthor{\bsnm{{Vernazza}}, \binits{J.E.}},
\bauthor{\bsnm{{Avrett}}, \binits{E.H.}},
\bauthor{\bsnm{{Loeser}}, \binits{R.}}:
\byear{1981},
\bjtitle{\apjs}
\bvolume{45},
\bfpage{635}.
\doiurl{10.1086/190731}.
\adsurl{1981ApJS...45..635V}.
\end{barticle}
\endbibitem

\bibitem[\protect\citeauthoryear{{Westendorp Plaza}
  \textit{et~al.}}{1998}]{1998ApJ...494..453W}
\begin{barticle}
\bauthor{\bsnm{{Westendorp Plaza}}, \binits{C.}},
\bauthor{\bsnm{{del Toro Iniesta}}, \binits{J.C.}},
\bauthor{\bsnm{{Ruiz Cobo}}, \binits{B.}},
\bauthor{\bsnm{{Martinez Pillet}}, \binits{V.}},
\bauthor{\bsnm{{Lites}}, \binits{B.W.}},
\bauthor{\bsnm{{Skumanich}}, \binits{A.}}:
\byear{1998},
\bjtitle{\apj}
\bvolume{494},
\bfpage{453}.
\doiurl{10.1086/305192}.
\adsurl{1998ApJ...494..453W}.
\end{barticle}
\endbibitem

\bibitem[\protect\citeauthoryear{{Westendorp Plaza}
  \textit{et~al.}}{2001}]{2001ApJ...547.1130W}
\begin{barticle}
\bauthor{\bsnm{{Westendorp Plaza}}, \binits{C.}},
\bauthor{\bsnm{{del Toro Iniesta}}, \binits{J.C.}},
\bauthor{\bsnm{{Ruiz Cobo}}, \binits{B.}},
\bauthor{\bsnm{{Mart{\'{\i}}nez Pillet}}, \binits{V.}},
\bauthor{\bsnm{{Lites}}, \binits{B.W.}},
\bauthor{\bsnm{{Skumanich}}, \binits{A.}}:
\byear{2001},
\bjtitle{\apj}
\bvolume{547},
\bfpage{1130}.
\doiurl{10.1086/318376}.
\adsurl{2001ApJ...547.1130W}.
\end{barticle}
\endbibitem

\bibitem[\protect\citeauthoryear{{Wu} and {Ai}}{1990}]{1990AcApS..10..371W}
\begin{barticle}
\bauthor{\bsnm{{Wu}}, \binits{L.-X.}},
\bauthor{\bsnm{{Ai}}, \binits{G.-X.}}:
\byear{1990},
\bjtitle{Acta Astrophys. Sinica}
\bvolume{10},
\bfpage{371}.
\adsurl{1990AcApS..10..371W}.
\end{barticle}
\endbibitem

\end{thebibliography}

\end{article}
\end{document}